\documentclass[sigconf]{acmart}
\AtBeginDocument{%
  \providecommand\BibTeX{{%
    \normalfont B\kern-0.5em{\scshape i\kern-0.25em b}\kern-0.8em\TeX}}}

\copyrightyear{2024}
\acmYear{2024}
\setcopyright{acmlicensed}\acmConference[ASSETS '24]{The 26th International ACM SIGACCESS Conference on Computers and Accessibility}{October 27--30, 2024}{St. John's, NL, Canada}
\acmBooktitle{The 26th International ACM SIGACCESS Conference on Computers and Accessibility (ASSETS '24), October 27--30, 2024, St. John's, NL, Canada}
\acmDOI{10.1145/3663548.3675615}
\acmISBN{979-8-4007-0677-6/24/10}

\acmConference[ASSETS'24]{ACM SIGACCESS Conference on Computers and Accessibility}{ October 28--30, 2024}{}
\acmBooktitle{ASSETS'24: ACM SIGACCESS Conference on Computers and Accessibility,
 October 28--30, 2024, St. John's, Newfoundland and Labrador} 
\acmISBN{978-1-4503-XXXX-X/18/06}

\acmSubmissionID{9438}
\usepackage{pdfcomment}
\usepackage{graphicx}
\usepackage{caption}
\usepackage{array}
\usepackage[export]{adjustbox}
\usepackage{float}
\usepackage{xcolor}
\usepackage{hyperref}

\begin{document}

\title{Our Stories, Our Data: Co-designing Visualizations with People with Intellectual and Developmental Disabilities}

\author{Keke Wu}
\affiliation{%
  \institution{University of North Carolina}
  \city{Chapel Hill}
  \country{United States}}
\email{kekewu@cs.unc.edu}

\author{Ghulam Jilani Quadri}
\affiliation{%
  \institution{University of Oklahoma}
  \city{Norman}
  \country{United States}}
\email{quadri@ou.edu}

\author{Arran Zeyu Wang}
\affiliation{%
  \institution{University of North Carolina}
  \city{Chapel Hill}
  \country{United States}}
\email{zeyuwang@cs.unc.edu}

\author{David Kwame Osei-Tutu}
\affiliation{%
  \institution{University of North Carolina}
  \city{Chapel Hill}
  \country{United States}}
\email{kwane622@email.unc.edu}

\author{Emma Petersen}
\affiliation{%
  \institution{University of Colorado}
  \city{Boulder}
  \country{United States}}
\email{emmajpetersen@gmail.com}

\author{Varsha Koushik}
\affiliation{%
  \institution{Colorado College}
  \city{Colorado Spring}
  \country{United States}}
\email{vkoushik@coloradocollege.edu}

\author{Danielle Albers Szafir}
\affiliation{%
  \institution{University of North Carolina}
  \city{Chapel Hill}
  \country{United States}}
\email{danielle.szafir@cs.unc.edu}

\begin{abstract}
Individuals with Intellectual and Developmental Disabilities (IDD) have unique needs and challenges when working with data. While visualization aims to make data more accessible to a broad audience, our understanding of how to design cognitively accessible visualizations remains limited. In this study, we engaged 20 participants with IDD as co-designers to explore how they approach and visualize data. Our preliminary investigation paired four participants as data pen-pals in a six-week online asynchronous participatory design workshop. In response to the observed conceptual, technological, and emotional struggles with data, we subsequently organized a two-day in-person co-design workshop with 16 participants to further understand relevant visualization authoring and sensemaking strategies. Reflecting on how participants engaged with and represented data, we propose two strategies for cognitively accessible data visualizations: transforming numbers into narratives and blending data design with everyday aesthetics. Our findings emphasize the importance of involving individuals with IDD in the design process, demonstrating their capacity for data analysis and expression, and underscoring the need for a narrative and tangible approach to accessible data visualization.
\end{abstract}




\begin{teaserfigure}
  \includegraphics[width=\textwidth]{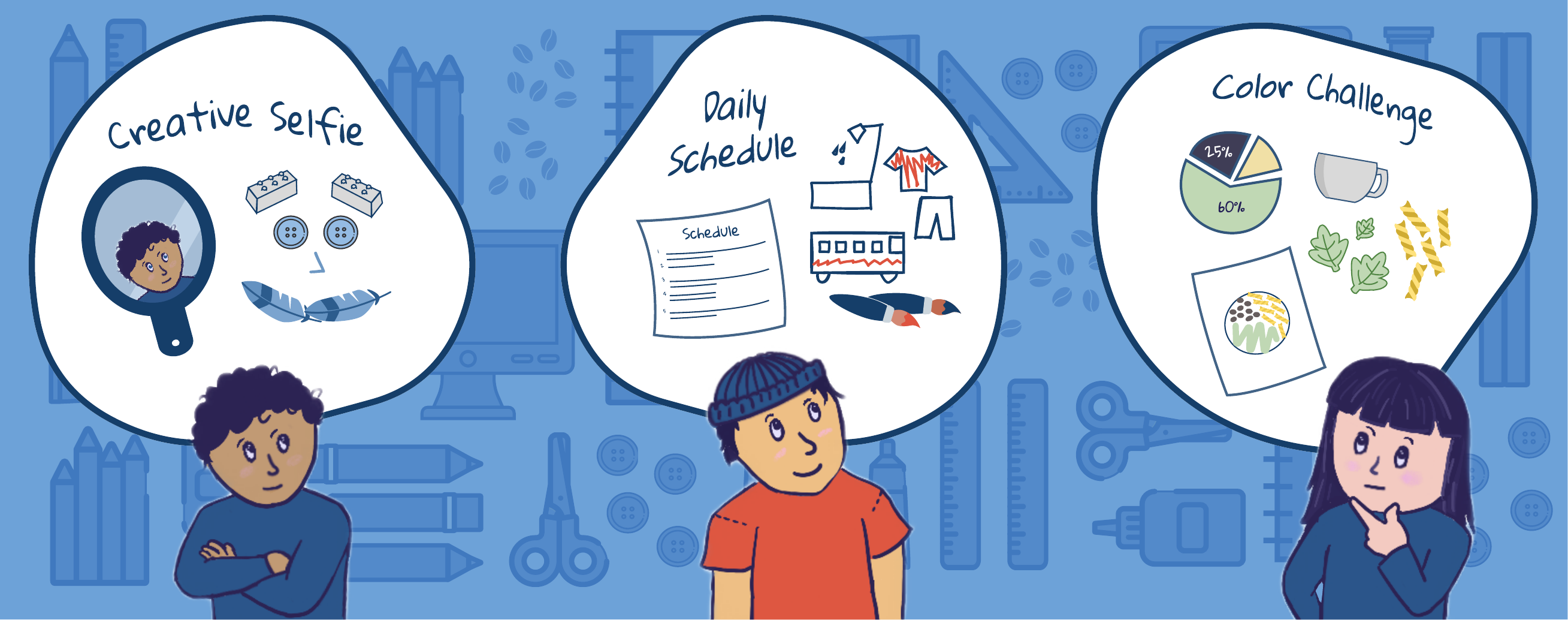}
  \Description{An overview of a data visualization co-design workshop. Three participants with IDD actively engage in the design process using tactile materials like LEGOs, beans, and pasta to create physical data representations.}
  \caption{Participants with IDD relied on personal experiences to interpret and contextualize data. To make data accessible, they engaged in hands-on activities, constructing their understanding of complex concepts. They humanized data with personal narratives, connecting abstract numbers with concrete human experiences. Additionally, they experimented with materiality and visual elements, using analogies and metaphors to create intuitive mappings that made the data more understandable.}
  \label{fig:teaser}
\end{teaserfigure}

\maketitle

\section{Introduction}
Intellectual and Developmental Disabilities (IDD) encompass a broad spectrum of conditions that significantly impact an individual's physical, intellectual, and/or emotional development trajectory \cite{about}. Common examples of IDD include Down syndrome, Fetal Alcohol syndrome, Fragile X syndrome, Autism spectrum disorder (ASD), and Cerebral Palsy \cite{criteria}. Each individual with IDD has a unique cognitive profile, characterized by varying degrees and combinations of challenges in intellectual functioning, such as learning, reasoning, and problem-solving, as well as daily adaptive behaviors, including communication, social interactions, and independent living \cite{criteria}. These challenges often hinder the effective use of conventional visualizations---such as charts, graphs, and maps---and result in distinct ways in which people with IDD interpret \cite{chi21, hdv}, conceptualize \cite{chi23}, and engage \cite{vis23} with data, requiring visualizations to be used and designed differently. Yet we lack guidance on how to address such differences effectively.

Due to impaired numerical understanding, reasoning, and abstract thinking \cite{about}, individuals with IDD tend to employ different strategies from the neurotypical population when interpreting visually encoded data \cite{chi21}. Furthermore, the unique life experiences of this population create innovative scenarios for data analytics and present new opportunities for visualization. These opportunities range from facilitating self-advocacy and storytelling for social inclusion to promoting self-expression and reflection for enhanced mental health to enabling self-regulation and management for everyday autonomy \cite{chi23}. As a cognitive technology, data visualization is designed to augment human cognition \cite{vision_think}, supposedly making data more accessible and comprehensible to a broader audience \cite{lee2020reaching}. However, conventional visualization practices typically assume a ``standard'' user and provide limited support in these scenarios, failing to capture diverse abilities and needs. This assertion means that guidelines and tools for visualization were often developed without considering individuals with IDD. Adhering strictly to traditional visualization approaches risks perpetuating this exclusion. In this paper, we collaborate with individuals with IDD to explore the role of visualization in enhancing data accessibility for their community. We define visualization in the broadest sense as graphical representations of data, designed to support both understanding and engagement. Our goal is to identify and establish best practices tailored to meet the distinct requirements of this population, generating cognitively inspired strategies and innovating the design of visualizations.

Co-design \cite{codesign}, a participatory approach that involves target users in defining problems and designing solutions, is an effective approach to engage this population in research~\cite{designwithid} and is critical for achieving this objective. It establishes a shared platform between researchers and participants, ensuring that the design of visualizations is not only relevant, effective, and accessible, but also reflective of the unique perspectives and needs of individuals with people with IDD \cite{cvo, marginalized, designwithid}. Additionally, co-design supports better community integration, enabling people with IDD to directly explore and analyze data impacting themselves and their communities \cite{voice}. This active involvement enables people with IDD to collaboratively address issues and improve their current condition. Finally, co-design situates data analytics within the cultural context of the community, facilitating a more comprehensive and culturally relevant approach to designing visualizations. This approach can ensure the visualizations resonate with the cultural norms, values, and preferences of the intended audience, fostering greater data engagement and understanding. In this study, we leverage co-design to unlock neurodiversity in data authoring, demonstrating how inclusive design practices can empower individuals with IDD to actively contribute to and communicate their unique perspectives through data-driven narratives.

We engaged 20 participants with IDD as active research partners and co-designers in two-phase hybrid workshops to explore how they approach and visualize data. Our exploration included a six-week preliminary phase conducted online and a comprehensive two-day in-person session. Drawing inspiration from the \textit{Dear Data} project \cite{deardata}, our first workshop was structured as a role-playing game. Four participants were paired as data pen-pals, exchanging visualizations following target prompts asynchronously over Zoom during the six-week period. In response to the observed conceptual, technical, and emotional struggles participants faced with data, we adapted the activities and formats accordingly. The second workshop, modeled after the \textit{Creative Visualization Opportunities (CVO) workshop} \cite{cvo}, was conducted as a two-day in-person synchronous event at a local community center. 16 participants with IDD were grouped into four cohorts based on shared interests: Art, Food, Games, and Sports. We gathered a diverse range of data to understand participants’ data sensemaking and visualization strategies, including design artifacts, recordings, interviews, observational notes, behavioral patterns, and reflective feedback.

Our findings revealed that individuals with IDD relied on personal experiences to interpret and contextualize data, infusing abstract data with personal meaning. Through hands-on exploration, personal narratives, and visual analogies, participants sought to create visualizations that resonate with their everyday activities and emotional experiences. Building on these insights, we propose alternative guidance for accessible visualization design, emphasizing two key strategies: first, transforming raw data into compelling narratives that bridge abstract numbers and human experiences; and second, blending data design with everyday aesthetics to create engaging and meaningful visualizations that enhance data appreciation. We advocate for a narrative and tangible approach to data visualization, reimagining its role in enhancing data accessibility for individuals with diverse abilities across data representation, expression, interaction, and education.

\noindent\textbf{Contributions:} 
The primary contribution of this work is the insights and solutions gained from co-designing data visualizations with 20 individuals with IDD, focusing on data sensemaking and authoring. We learned that participants often used personal experiences to interpret and contextualize data. Specifically, they employed hands-on activities to explore data, humanized data with personal narratives, and made data more understandable through visual analogies and metaphors. These findings reveal that integrating personal experiences and familiar contexts into data visualization can significantly enhance understanding and engagement for individuals with IDD. Based on these insights, we propose two strategies for cognitively accessible data visualizations: first, transforming raw data into compelling narratives to bridge the gap between complex data and relatable human experiences; and second, integrating data design with everyday aesthetics to develop visualizations that resonate with and are accessible to diverse audiences. We advocate for a narrative and tangible approach to data visualization, promoting inclusive data access across diverse abilities.

\section{Background}
We build on past studies utilizing co-design for cognitive accessibility and visualization, examine inclusive data communication and representation techniques, and explore storytelling's role in supporting people with IDD. 

\subsection{Co-design for Cognitive Accessibility and Visualization}
Co-design~\cite{spinuzzi2005methodology} integrates user experiences and needs directly into the design process, ensuring solutions are not only functional but also aligned with their intended uses. Co-designing with individuals with cognitive disabilities presents a range of unique opportunities and challenges~\cite{sarmiento2015co,wimer2024beyond}. Past research has discussed barriers~\cite{hendriks2015codesign}, strategies~\cite{colin2020lessons}, adaptations~\cite{respectful}, and applications~\cite{heerings2022ask, how2017envisioning, koushik2022towards, rathnayake2021co} of co-design for this population. For instance, co-design activities often require cognitive tasks such as abstraction and conceptualization, which may hinder effective participation for individuals with cognitive disabilities. Employing scenario-based design methods~\cite{koushik2022towards} and tangible objects~\cite{colin2020lessons} can enhance involvement and understanding, overcoming comprehension and digital exclusion barriers. Moreover, cognitive disabilities frequently limit communication abilities, presenting challenges for participants in effectively expressing their perspectives and ideas. Leveraging diverse modalities, including visual aids, storyboards, interviews, and prototypes, can help elicit the needs of participants~\cite{kim2024opportunities, colin2020lessons}. Additionally, integrating proxies or experts can facilitate communication and ensure meaningful participation~\cite{koushik2022towards, colin2020lessons}. Adopting competency-based design approaches~\cite{bayor2019howtoapp, koushik2022towards} will ensure that activities are personally relevant, appropriately adapted, and supportive of self-expression, determination, and autonomy.

In data visualization research, co-design has proven effective in developing meaningful solutions that address diverse user needs~\cite{cvo, chartreader} and bridge the gap between domain expertise and practical application~\cite{visgap}. For example, Khowaja et al.\cite{activis} collaborated with local healthcare professionals to co-design ActiVis, a tool for monitoring patient physical activity through wearable sensors. Snyder et al.\cite{cancer} involved prostate cancer survivors with limited graph literacy in co-designing timeline visualizations to improve health tracking. León et al.\cite{co-creation} evaluated co-design with social science researchers, emphasizing the importance of perceived direct benefits and personal relevance. Lundgard et al.\cite{lundgard2019sociotechnical} worked with blind users to address sociotechnical considerations for accessible visualizations, underscoring the value of equal partnership with people with disabilities. These examples illustrate how involving end-users in the design process leads to practical and impactful solutions.

Co-design offers an intuitive means for understanding the needs and challenges of people with IDD when interacting with data. Notably, individuals with IDD engage with data differently~\cite{chi21}, and the abstraction of data employed in common visualization techniques can lead to anxiety and disconnection~\cite{chi23}. 
Our workshop builds on co-design traditions from both accessibility~\cite{designwithid} and visualization~\cite{cvo}. Specifically, we
(1) tailor data-related activities to the personal experiences of participants to make data more personal and approachable~\cite{Neate2019, Acton2023}, (2) encourage hands-on and constructionist approaches~\cite{constructive, constructive_token} to make abstract concepts more tangible and understandable, and (3) actively involve caregivers in the process to gain a deep and authentic understanding of neurodivergent experiences and perspectives~\cite{koushik2022towards, proxy}.

\subsection{Inclusive Data Communication and Representation}
Inclusive data communication and representation are crucial for democratizing access to information, ensuring that everyone, regardless of their educational backgrounds, abilities, or sensory processing needs, can benefit from data-driven insights~\cite{lee2024inclusive}. This entails designing data delivery systems and visualizations that are accessible to a diverse audience, enabling a broader range of people to make informed decisions based on the data presented. Previous studies have characterized how various disabilities affect interactions with data~\cite{marriott2021inclusive, lee2024inclusive, wimer2024beyond}, emphasizing the need for designs that accommodate diverse abilities while ensuring clear benefits for individuals or the broader community. Further studies have probed into multisensory, tangible aspects of data interaction to engage novices in working with data~\cite{voxlens, chundury2021towards, bae2022making, bae2022cultivating}, aiming to enhance the usability and accessibility of data across different user groups and modalities. Additionally, studies have examined the role of personal relevance~\cite{dataispersonal,datahunches} and the use of visualization as a rhetorical device~\cite{flack2019lego, self-reflection, mementos} to strengthen the connection between individuals and data, tailoring visualizations to serve a variety of personal and practical purposes. These efforts collectively underscore the vital role of inclusivity in effective data communication, ensuring that data serves as a powerful and accessible tool for diverse users in various contexts.

Cognitive and learning disabilities present significant challenges in how data and visualizations are consumed and created, often rendering conventional tools ineffective for those with these disabilities. Past studies showed that individuals with IDD were particularly sensitive to the design of visualizations, suggesting that cognitively accessible visualizations might require new guidelines~\cite{viscomm, chi21}. Further research highlighted the need for human-centered approaches to data visualization that address their specific needs~\cite{chi23}, emphasizing the importance of making data personal and approachable~\cite{vis23}. These approaches prioritized scenarios such as telling stories with personal data for effective self-advocacy and social participation, tracking and synthesizing self-assembled datasets for expression and meaning-making, and enhancing data agency and autonomy by developing intuitive data representations and interactions that support self-regulation and independence.

Yet, we currently lack a comprehensive understanding of and accommodations for the unique needs and challenges faced by individuals with IDD when interacting with data and creating visualizations. Moreover, there is a significant gap in inclusion and representation within the design process, with individuals with IDD historically marginalized in visualization development. With this in mind, our workshop focuses on three key strategies to enhance cognitive accessibility in data communication and representation:
(1) We employ a constructionist approach~\cite{huron2014constructive}, advocating for the use of everyday materials to make data more relatable and tangible. (2) We combine self-discovery and data representation to deepen participants' connection to and understanding of their own behaviors, preferences, and conditions. (3) We ground data-driven activities in everyday contexts, making the data and its implications directly relevant to participants' daily lives.

\subsection{Supporting People with IDD through Data Storytelling}
Storytelling offers significant benefits in supporting individuals with IDD. Research has demonstrated its therapeutic effects in bolstering self-esteem~\cite{therapeutic} and addressing sensory issues~\cite{sensorystory} among children with learning disabilities. It has also been shown to enhance cognitive functions such as memory, attention, and concentration~\cite{Rahgoi}, and its educational and ethical implications include fostering learning and empathy among people with Autism~\cite{digital, socialempathy}. By providing a platform for individuals with IDD to express themselves creatively, storytelling helps discern their strengths, preferences, and abilities, empowering them to participate in research and decision-making~\cite{saridaki2018digital, decision, learnandapply}.

Narrative visualization, a subset of visualization research dedicated to storytelling with data~\cite{segel2010narrative}, has gained considerable popularity in recent years. Progress has been made in enhancing data story authoring~\cite{lee2013sketchstory, iStory}, improving annotation and interaction~\cite{chartaccent}, 
and in diversifying its design and display~\cite{datacomics, character, storytellingtv}. Additionally, research has evaluated the effectiveness of data storytelling in supporting cognition~\cite{recall}, engagement~\cite{engagement}, and its influence on empathy and attitude~\cite{attitude}. 
Recognizing its pivotal role in supporting public participation, researchers have started to explore the link between data storytelling and social good. For example, Erete et al.~\cite{npo} interviewed non-profit organizations (NPOs) and found that data narratives were often used to solicit support from grantors and other stakeholders. Ortega et al.~\cite{personaldatacomics} investigated the impact of involving non-experts in the creation of personal data comics as a means to support data literacy and agency. Sakamoto et al.~\cite{affectstorytelling} empirically explored the role of affect in a data video to convey data-driven messages and found its positive influence on viewers' perception. Lund~\cite{hiphop} discussed the importance of bringing in a social justice mindset to data storytelling to compel readers and viewers to act upon findings. 

These examples collectively illustrate the potential of storytelling in supporting accessible data interactions for individuals with IDD, who face unique challenges when working with data and often require different approaches. In our workshop, storytelling serves multiple purposes: (1) We embrace a narrative workshop format to enhance engagement, comprehension, and foster positive data experiences. (2) We empower individuals with IDD as storytellers, employing relatable prompts and personally relevant datasets to encourage creative self-expression through visualization. (3) We harness visual storytelling as a means to cultivate a shared identity, enabling collective sensemaking~\cite{collective}, social inclusion, and deriving culturally responsive solutions from the community to inspire the design of more accessible visualizations.

\section{Methods}

We conducted two hybrid co-design workshops involving 20 participants with IDD to explore their approaches to data visualization. The initial workshop paired four participants as \textit{data pen-pals}, facilitating asynchronous exchanges of visualizations over a six-week period via Zoom. Due to observed technical and emotional challenges, along with a high dropout rate, the second workshop was restructured to encourage more direct social interactions and reduce technological barriers for participation. In collaboration with \textit{a local community center}, we organized 16 participants into four cohorts based on shared interests and conducted the workshop as a two-day in-person event.

Our workshop activities drew inspiration from prior research on data accessibility needs for individuals with IDD~\cite{chi21, chi23}. These activities were developed with input from IDD self-advocates and domain experts. Participants engaged in various data-related activities, including recalling and collecting personal data, analyzing and visualizing provided datasets, and redesigning existing visualizations (e.g., pie charts). The data we collected included design artifacts, photographs, videos, screen recordings, surveys, field notes, and interviews with participants, their caregivers, and professionals from the community center. To analyze this data, we applied thematic analysis~\cite{braun2012thematic} and open coding techniques~\cite{glaser1967discovery} to identify patterns and extract insights.

\begin{figure*}[t]
\centering
\includegraphics[width=\textwidth]{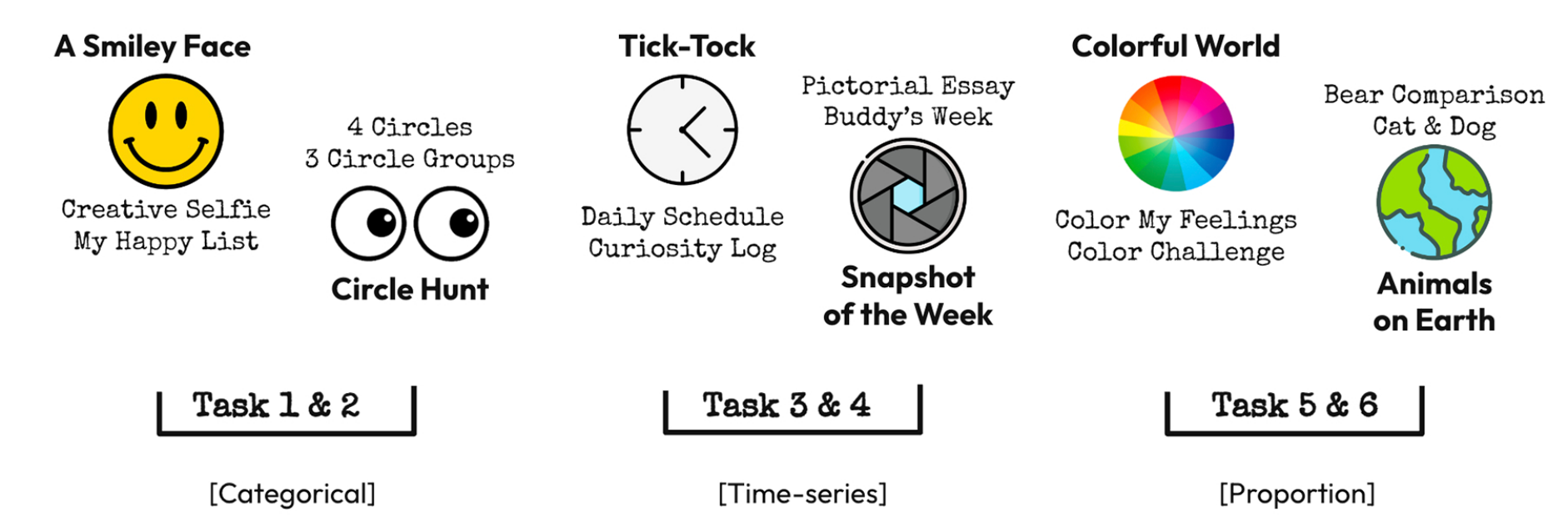}
\Description{An overview of the workshop "Aliens from the VisuaLand" including 6 tasks, focusing on categorical data, time-series data, and proportion data.}
\centering
\caption{Overview of the workshop \textit{Aliens from the VisuaLand} (``The Circle Story''). Our workshop comprised 6 tasks exploring three types of data. Tasks 1\& 2 centered on how people with IDD visually depict data pertaining to themselves and their surroundings, particularly focusing on categorical data. Tasks 3 \& 4 examined methods they employ to represent and manage time effectively, emphasizing time-series data. Tasks 5 \& 6 were dedicated to integrating individuals into communities and uncovering their approaches to visualizing part-to-whole relationships within the data.} 
\label{fig:activity}
\end{figure*}

\subsection{Study Design}
Prior studies on people with IDD note that the idea of ``data'' may often be an intimidating and uncomfortable concept \cite{chi23}. To make data more approachable and the activities enjoyable, we fictionalized the workshop's setting with an alien narrative. Four alien characters were designed to represent four topics of interest (i.e., Arts, Sports, Games, and Food). In this narrative, the aliens were from an imaginary planet ``the VisuaLand,'' and an unknown force destroyed their home. The participants were, therefore, invited to create visualizations aimed at helping the aliens understand life on Earth, inviting the aliens to consider relocation. 

To encourage creative self-expression, we employed open-ended prompts and personally relevant datasets throughout the workshop. Across six design activities exploring categorical, time-series, and proportion data (illustrated in Figure \ref{fig:activity}), participants were invited to craft visualizations showcasing aspects of themselves, their experiences, and their communities. To minimize the Hawthorne Effect~\cite{McCambridge2014}, we communicated the study’s purpose beforehand, assured participants of confidentiality and established rapport through a series of ice-breaker activities. We provided multiple examples for each activity to illustrate the diverse range of possibilities, encouraging authentic expression and reducing the influence of observation. The activities were iteratively designed over the course of two years in collaboration with IDD self-advocates and domain experts to meet participants where they are while also probing their mental models to understand how they construct visual representations for various data types. Detailed task descriptions and workshop slide decks can be found at \href{https://osf.io/dsvn8/?view_only=16497f3c27dc4f78ba0cc79a87ecad75}{\textbf{{IDD Co-design Workshop}}}.
Below, we summarize each task, its design objectives, and the data involved, highlighting any modifications made for in-person participants.

\begin{figure*}[t]
\centering
\includegraphics[width=\textwidth]{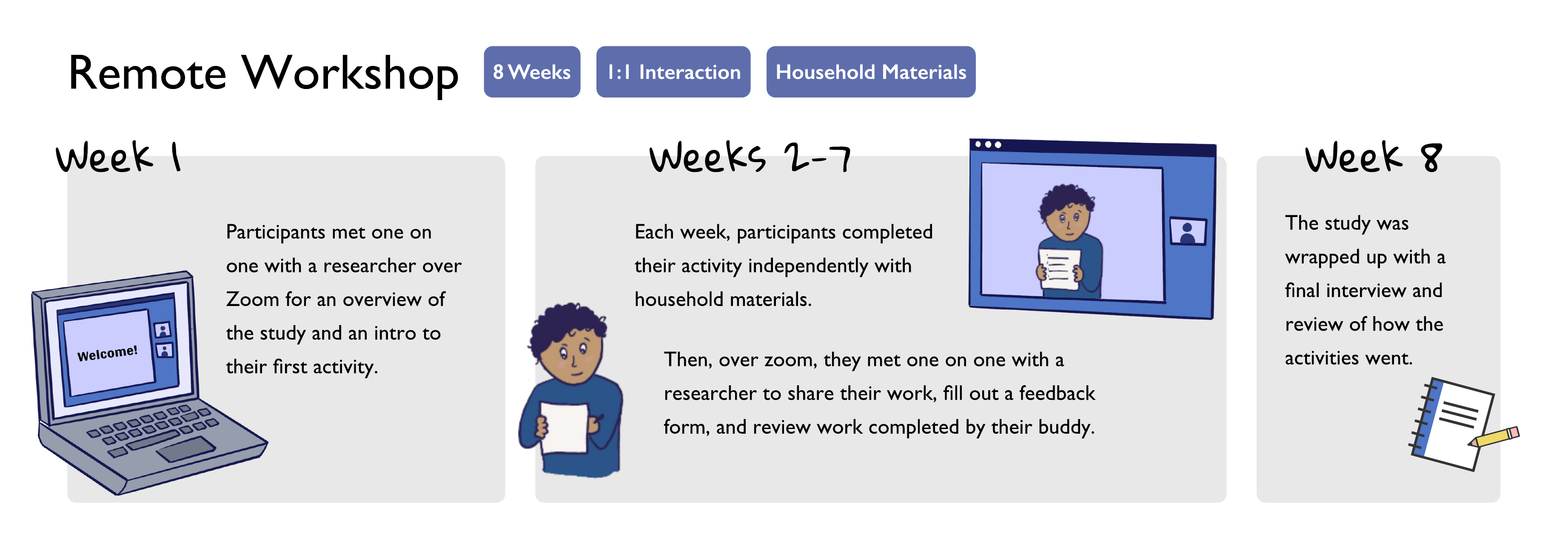}
\Description{An illustration of the 8-week online workshop data pen-pal system}
\caption{Overview of the online workshop structure. The online segment spanned eight weeks, during which four participants paired up as data pen-pals. Each week, they met individually with a researcher via Zoom. These sessions involved sharing their own work, reviewing their buddy's work, and previewing the upcoming week's activity. Materials for authoring the visualizations were not provided.}
\label{fig:remote}
\end{figure*}

\section*{Categorical Data}

\begin{enumerate}
\item\textit{Creative Selfie}
    \begin{enumerate}
        \item Prompt:
        Create a selfie not taken by a phone or camera.
        \item Objectives:
        Encourage creative self-expression and novel ways of visually representing the self. Get participants familiar with creating visual representations.
    \end{enumerate}
    
    \item \textit{My Happy List}
    \begin{enumerate}
        \item Prompt:
        Make a visually ranked happy list that contains five things making you happy.
        \item Objectives:
        Promote self-awareness, gratitude, and well-being. Encourage and reinforce the understanding of one's emotions, preferences, and values. Foster a positive orientation towards the concept of ``data.''
    \end{enumerate}

    \item \textit{Four Circles}
    \begin{enumerate}
        \item Prompt:
        Find four circular objects related to your favorite topic from Arts, Sports, Game, and Food.
        \item Objectives:
        Encourage participants to draw connections between concepts and objects. Work with patterns in data's form (i.e., similarity and difference).
    \end{enumerate}

   \item \textit{Three Circle Groups}
    \begin{enumerate}
        \item Prompt:
        Create three circle pairs according to your criteria.
        \item Objectives:
        Engage participants in a creative and analytical process of selecting and pairing circles. Understand basic categorical mapping paradigms.
    \end{enumerate}    
\end{enumerate}

\section*{Time-series Data}

\begin{enumerate}
    \item \textit{Daily Schedule}
    \begin{enumerate}
        \item Prompt:
        Create a visualization of your daily schedule.
        \item Objectives:
        Represent temporal event sequences. Understand intuitive mappings for categorical events and time.
    \end{enumerate}
    
    \item \textit{Curiosity Log}
    \begin{enumerate}
        \item Prompt:
        Pick one day, observe yourself, and take notes whenever you are curious about something. Create a visual log of this process.
        \item Objectives:
        Understand information-seeking behaviours, and promote self-awareness, reflection, and learning.
        \item Modifications:
        Simplified to create a visual log of one curious thing at the moment.
    \end{enumerate}

    \item \textit{Pictorial Essay}
    \begin{enumerate}
        \item Prompt:
        Create a pictorial essay that has at least five images to highlight your week. 
        \item Objectives:
        Allow participants to capture memorable moments and tell a compelling story about their week and their unique experiences using this data.  
        \item Modifications:
        Provided sticky notes to scaffold the task.
    \end{enumerate}

   \item \textit{Buddy's Week}
    \begin{enumerate}
        \item Prompt:
        Read the data tracker table of your alien buddy's activities and try creating a visualization to communicate your buddy's week.
        \item Objectives:
        Understand how participants interpret data from a table and represent time-series data. Use visualizations to summarize an event sequence.
        \item Modifications:
        Simplified the table from three to one row.
    \end{enumerate}    
\end{enumerate}

\section*{Proportion Data}

\begin{enumerate}
    \item \textit{Color My Feelings}
    \begin{enumerate}
        \item Prompt:
        Finish the three-step color my feelings activity: 1) fill nine emoji with different colors, 2) find nine objects of the color you selected for ``happy'', 3) rate the nine objects from 1--9 based on their colors.
        \item Objectives:
        Explore how participants use colors to communicate feelings and emotions, and understand their color concept associations.
        \item Modifications:
        Broke the activity down into steps.
    \end{enumerate}
    
    \item \textit{Color Challenge}
    \begin{enumerate}
        \item Prompt:
        Create a visualization to represent your buddy's diet, specifically 10\% of fat, 60\% of veggies, and 30\% of grains \& starches.
        \item Objectives:
        Understand accessible representations of percentages and proportion data, especially in light of known accessibility limitations of pie charts \cite{chi21}.
    \end{enumerate}

    \item \textit{Bear Comparison}
    \begin{enumerate}
        \item Prompt:
        Read a provided table (containing different metrics about the diet, skills, and rareness of brown bears, polar bears, and panda bears), and create a visualization for comparing three popular bear species on Earth.
        \item Objectives:
        Understand how participants interpret and compare multivariate data and represent proportion.
        \item Modifications:
        Removed for in-person participants.
    \end{enumerate}

   \item \textit{Cats \& Dogs}
    \begin{enumerate}
        \item Prompt:
        Use provided information (a series of visualizations containing a map comparing the popularity of both animals by state, different charts and graphs illustrating their social media presence, and owners' ratings on different metrics), and create a visual story about cats \& dogs.
        \item Objectives:
        Understand how participants interpret various visualizations (e.g., maps, bars, treemaps, line charts, and isotype visualizations~\cite{isotype}) and represent this information. Understand representations across multiple elements for data storytelling.
        \item Modifications:
        Simplified to show only owners' rating.
    \end{enumerate}    
\end{enumerate}

These tasks emphasized either personal data to connect to concepts of personal relevance and make the experience more meaningful~\cite{kadijevich_angeli_schulte_2013, broadening} or simple, playful datasets to minimize potential disengagement from negative stereotypes with data and mathematical literacy often faced by people with IDD \cite{about, chi23}. Additionally, tasks were structured to gradually increase in complexity throughout the series of activities, providing a steady scaffold to help participants develop a more sophisticated understanding of data.

\begin{figure*}[t]
\centering
\includegraphics[width=\textwidth]{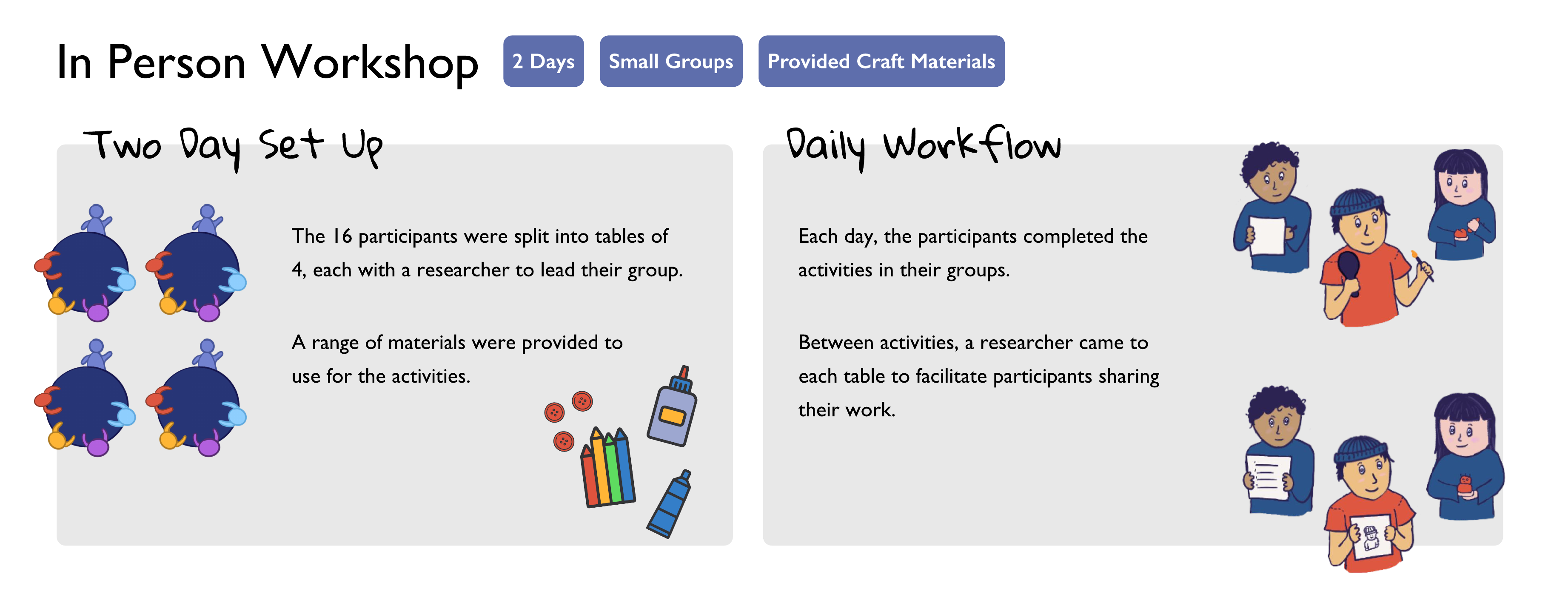}
\Description{An illustration of the two-day in-person workshop layout. 16 participants were put into groups of 4.}
\caption{Overview of the in-person workshop structure. In the in-person session, 16 participants were grouped into tables of four, each accompanied by a researcher. Across two days, each table engaged in three activities daily, exchanging feedback on their work after each task. Materials provided included craft supplies, food items, and constructive materials such as LEGOs.}
\label{fig:inperson}
\end{figure*}

\subsection{Online Data Pen-Pal Workshop}
Inspired by the Dear Data project~\cite{deardata}, our preliminary workshop paired four online participants with IDD as data pen-pals to create visualizations using provided prompts and facilitate visual communication in the VisuaLand structured role-playing scenario described above. Participants were paired randomly with a pen-pal and instructed that they would be communicating using the outputs of each activity. They attended weekly individual Zoom sessions, each lasting about 60 minutes and facilitated by a designated ``visual messenger'' from our research team. This researcher assisted in exchanging and discussing their visualizations, ensuring a smooth and effective communication process throughout the project. Each participant committed to an eight-week workshop schedule, which included a one-hour pre-workshop interview, six weeks of activity reviews and critiques, and a one-hour post-workshop interview. This setup is illustrated in Figure \ref{fig:remote}.

Prior to the workshop, participants engaged in a pre-interview Zoom call where they discussed their data and visualization experiences. During this session, they completed a visualization questionnaire comprising 10 questions to gauge baseline visualization literacy adapted from a popular data visualization literacy assessment test \cite{lee_vlat_2017}. Additionally, participants finished a practice screen-time visualization task to get a sense of the activity types and provided feedback. Finally, they were introduced to Task One (i.e., Creative Selfie). We walked them through a series of example slides, encouraging the use of a broad range of materials for their projects. Following the introduction, each participant received access to a personal Google Drive folder containing the activity slides for the workshop, broken down by activity.

Over the next six weeks, participants engaged in weekly Zoom calls, approximately one hour each, with their visual messenger. During these sessions, they discussed their own visualizations for the week and provided feedback on those created by their design buddy-—a fellow workshop participant—-using a pre-designed feedback report (accessible at \href{https://osf.io/dsvn8/?view_only=16497f3c27dc4f78ba0cc79a87ecad75}{\textbf{{IDD Co-design Workshop}}}). Discussions and comments centered on three key themes: engagement with the activity, comprehension of each other’s work, and suggestions for improvement. To maintain confidentiality and focus on visual communication, participants remained anonymous to one another throughout the workshop.

In post-workshop interviews, participants engaged in a comprehensive review and reflection of all activities and visualizations, both their own and those of their peers. They focused on the lessons learned from the workshop, proposed changes they would like to implement, and discussed their attitudes toward data and visualization. Upon completing the workshop, each participant received a \$120 Amazon Gift Card.

The remote setup provided participants the flexibility to create visualizations using readily available materials, which accommodated diverse schedules and preferences. However, this arrangement also introduced several challenges. The extended duration of engagement over six weeks 
in some cases led to fatigue and dropout. Additionally, the asynchronous 
communication between participants and the research team could delay immediate engagement and timely resolution of questions or concerns. Furthermore, conceptual challenges and technological literacy issues, such as understanding tasks and uploading visualizations to Drive, posed additional obstacles for participants. Notably, while participants were generally engaged with the activities, three out of seven participants withdrew from the study at various stages due to the significant difficulty in understanding data—-especially numeric data-—or the time commitment required.

\subsection{In-Person Co-Design Workshop}
Drawing from insights gained from the online workshop, we refined the workshop flow and activities into a two-day in-person co-design workshop at a local community center with 16 participants with IDD. Prior to the workshop, participants were divided into four cohorts 
based on their interests (Art, Food, Games, and Sports) with assistance from the community center staff. These topics were used to narrow down the design prompts and connect them to participants' interests. This setup is illustrated in Figure \ref{fig:inperson}. For the in-person workshop, we implemented several modifications from the remote study to enhance the experience: 

\noindent\textbf{(1) Data Simplification and Prompt Adjustments:} After Day One, the team reflected on the workshop experience and re-evaluated the difficulty level of Day Two's activities for participants. Based on feedback from the center staff, we simplified some datasets and adjusted the prompts. For instance, in the Buddy’s Week activity, we reduced the data table from three rows to one and supplied each participant with five sticky notes for creating their visualization. We also removed the Bear Comparison activity and streamlined the Cats \& Dogs activity to focus solely on comparing owners’ ratings.

\noindent\textbf{(2) Introduction of Templates and Scaffolding:} We observed significant cognitive challenges and even emotional burnout experienced by in-person participants when working with numerical data and concepts such as percentages. To support the participants more effectively, we introduced templates and added scaffolding measures. For example, in the Color My Feelings project, we divided the activity into three steps, progressing to the next only after all participants had completed the current one. In the Color Challenge activity, participants were provided with either a plate or a popsicle stick to use as a base for their visualizations, rather than starting from scratch. This approach was employed to evaluate the effectiveness of various instructional interventions.

\noindent\textbf{(3) Enhanced Interaction and Collaboration:} We supplied a variety of crafting materials and arranged participants into four groups at separate tables. Four facilitators from our team, one at each table, led three to four participants and collaborated closely with the center staff to clarify the prompts and offer support. This setup fostered a more interactive and collaborative environment, aiming to make the workshop more engaging for everyone involved.

Each table was equipped with identical supplies, comprising colored pencils, paper, wood tiles and blocks, play-doh, LEGO bricks, stickers, sticky notes, plates, and popsicle sticks. Extra materials such as pasta, beans, marbles, feathers, pom-poms, and buttons were arranged on a separate table. One facilitator from the research team was assigned to manage each table, collaborating closely with participants and their caregivers over the two days. We followed the center's operating schedule and had a total of four sessions over the course of two days. 

During the opening icebreaker, participants introduced themselves and shared their chosen topics. For each activity, we provided a task explanation and presented several examples. Subsequently, participants broke into smaller groups to work on their visualizations. Facilitators' primary responsibilities included clarifying tasks to their respective tables, ensuring everyone comprehended the prompts, aiding participants in the ideation or execution of their visualizations when necessary, gathering feedback on activities from participants, and taking observational notes. Each participant received a \$200 Walmart Gift Card as compensation for their participation.

The in-person setup proved to be more effective in engaging participants with IDD in co-design activities. This format highlighted the importance of multimodal and hands-on learning approaches in accessible data literacy. Providing tangible materials and step-by-step scaffolding helped participants better understand and engage with complex data concepts. It also encouraged spontaneous collaboration and peer support among participants, which was less feasible in a remote setting. Overall, the in-person format created a more immersive and supportive environment that significantly improved engagement and outcomes in the co-design activities.

\subsection{Participants}
We obtained approval from the institution’s Institutional Review Board (IRB) office before starting the recruitment process. Our workshops consisted of 20 participants aged 23 to 72, evenly divided between 10 females and 10 males, all with self-reported mild to moderate IDD. 
Participants were recruited based on functioning abilities such as being self-directed and capable of understanding and communicating, rather than through cognitive assessments (e.g., IQ tests). 

The four online participants were returning participants who had participated in previous studies, while the 16 in-person participants were recruited at a local community center, with no previous engagement in research. %
Participants’ de-identified demographic information is summarized in Table \ref{tab:participants}. 
We do not link artifacts or quotes with participant demographics per our partner organization's policies. Online participants tended to be younger, more verbal, and independent, with more experience in working with data and visualization. In contrast, in-person participants were typically older, less verbal, and less independent, with limited exposure to data and visualization. However, these differences did not significantly impact 
our results. 

\begin{table}[htbp]
    \centering
    \begin{tabular}{cccc}
    \toprule
    \textbf{Group} & \textbf{Age} & \textbf{Gender} & \textbf{IDD Type} \\
    \midrule
    Online & 23 & M & Cerebral Palsy\\
    Online & 30 & M & Autism\\
    Online & 36 & M & Autism\\
    Online & 41 & F & Mild ID\\
    In-person & 37 & M & Mild ID, Autism\\
    In-person & 39 & M & Moderate ID, Sturge-Weber syndrome\\
    In-person & 42 & M & Mild ID\\
    In-person & 53 & F & Mild ID, Sandhoff disease, Seizure\\
    In-person & 52 & F & Cerebral Palsy\\
    In-person & 68 & F & Mild ID\\
    In-person & 69 & M & Moderate ID\\
    In-person & 47 & F & Moderate ID\\
    In-person & 54 & M & Mild ID\\
    In-person & 34 & M & Down syndrome\\
    In-person & 36 & F & Down syndrome\\
    In-person & 32 & F & Mild ID\\
    In-person & 36 & M & Mild ID, Autism\\
    In-person & 72 & F & Moderate ID, Autism\\
    In-person & 64 & F & Mild ID\\
    In-person & 37 & F & Moderate ID\\
    \bottomrule
    \end{tabular}
    \caption{Participants' de-identified demographic information.}
    \label{tab:participants}
\end{table}

\subsection{Analysis}
For both the online and in-person workshops, we collected design artifacts created by participants, supplemented by our own observational notes. We also recorded video footage of the online sessions and took photographs during the in-person workshops to aid further analysis. We employed thematic analysis and open coding to identify patterns within the data. 
Our analysis focused on the unique strategies and challenges faced by participants in data interaction, visualization design, and the practical implications for engaging them in both online and in-person co-design settings.

Our team reviewed all artifacts collectively and conducted two rounds of discussions, structured by both activity and participant. These discussions helped us develop and refine a comprehensive set of 
codes. Throughout the discussions, we examined the strategies and challenges participants encountered in data comprehension and representation for each activity. We also assessed individual strengths and weaknesses, observed developmental changes over time, and recognized the dynamic role of visualization in participants' interactions with data. The complete collection of labeled artifacts is available at~\href{https://osf.io/dsvn8/?view_only=16497f3c27dc4f78ba0cc79a87ecad75}{\textbf{{IDD Co-design Workshop}}}.

\begin{figure*}
\includegraphics[width=\textwidth]{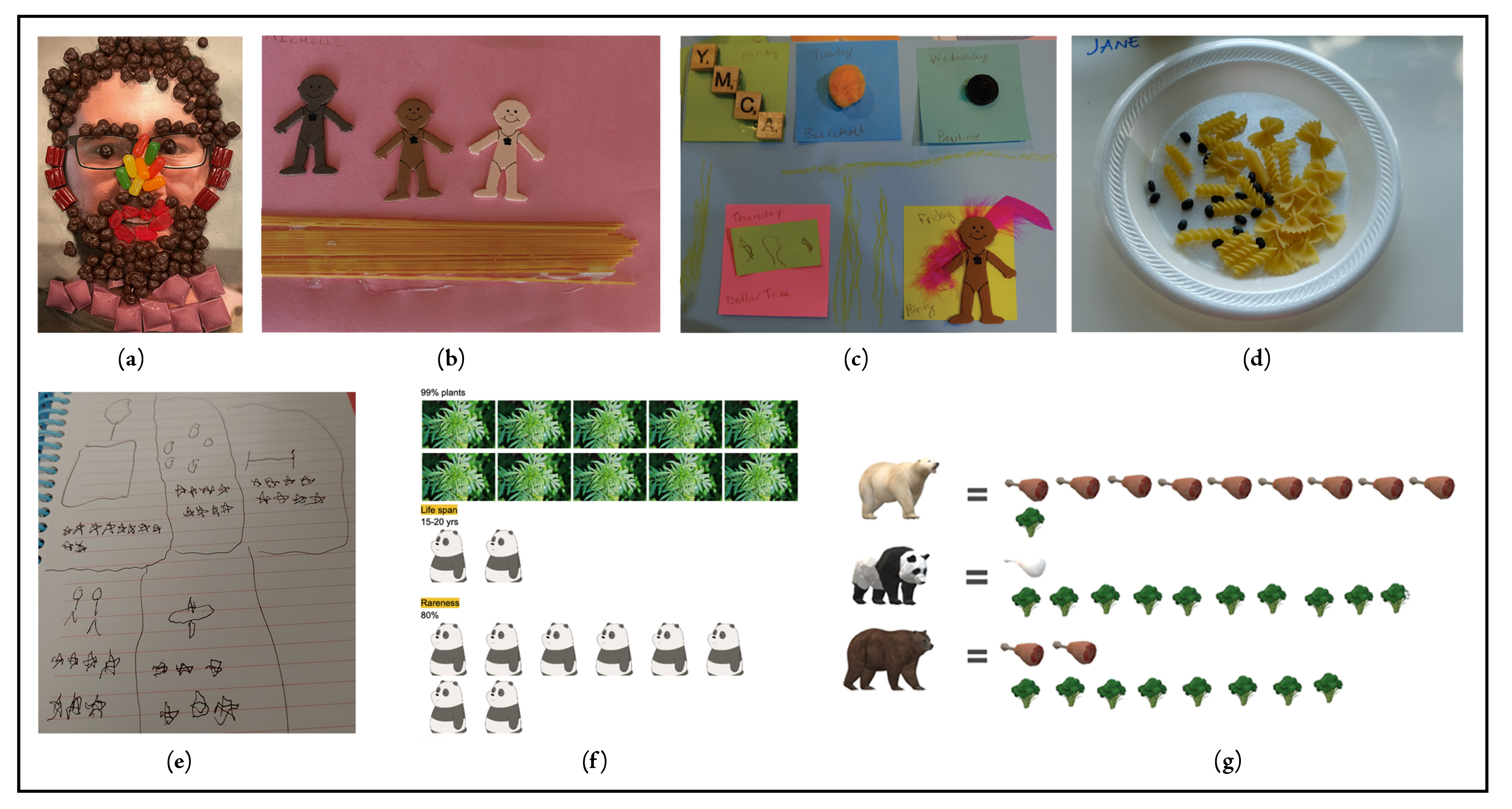}
\Description{A collage of artifacts made by participants, showing how they use various materials to create tangible data visualizations.}
\centering
\caption{Exploring data through hands-on activities. Participants in the workshop embraced a constructionist approach to data visualization, employing diverse materials to bring the data to life, and utilized isotype visualizations (e, f, g) to make data more concrete. Participants used candies to represent facial features (a) and utilized stick figures and pasta to symbolize the bond between themselves and loved ones (b). They crafted tangible pictorial essays (c) and favorite activity representations (d), using everyday objects to convey their experiences and emotions.} 
\label{fig:constructionist}
\end{figure*}

\begin{figure*}
\includegraphics[width=\textwidth]{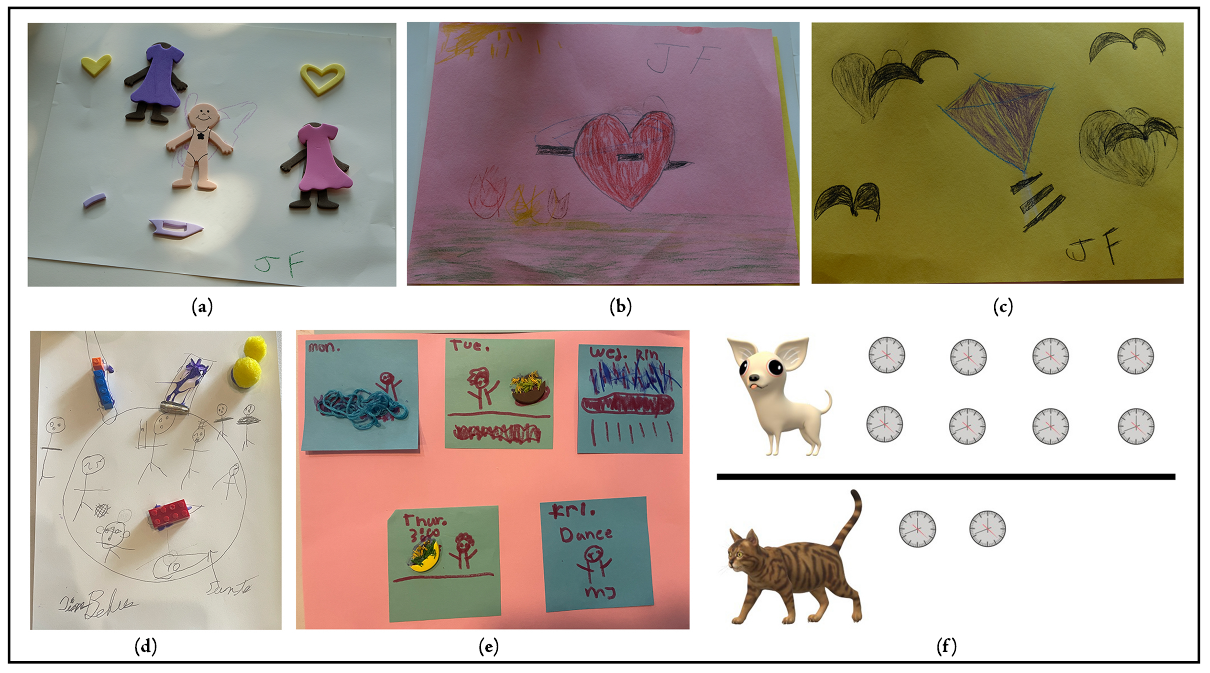}
\Description{A collage of artifacts made by participants, showing how they humanize data through personal narratives.}
\centering
\caption{Humanizing data through personal narratives. Participants were trying to attribute personal significance to the data, connecting it with their own lived experiences. They frequently integrated personally meaningful symbols, such as a heart, into their visualizations (a, b, c). They crafted characters and envisioned a complete scene as part of their Creative Selfie (d). They also developed a tangible narrative full of details depicting their week in the Pictorial Essay (e). They associated data points with emotional moments, capturing these sentiments in their representation of the Cats \& Dogs comparison (f).} 
\label{fig:humanize}
\end{figure*}

\begin{figure*}
\includegraphics[width=\textwidth]{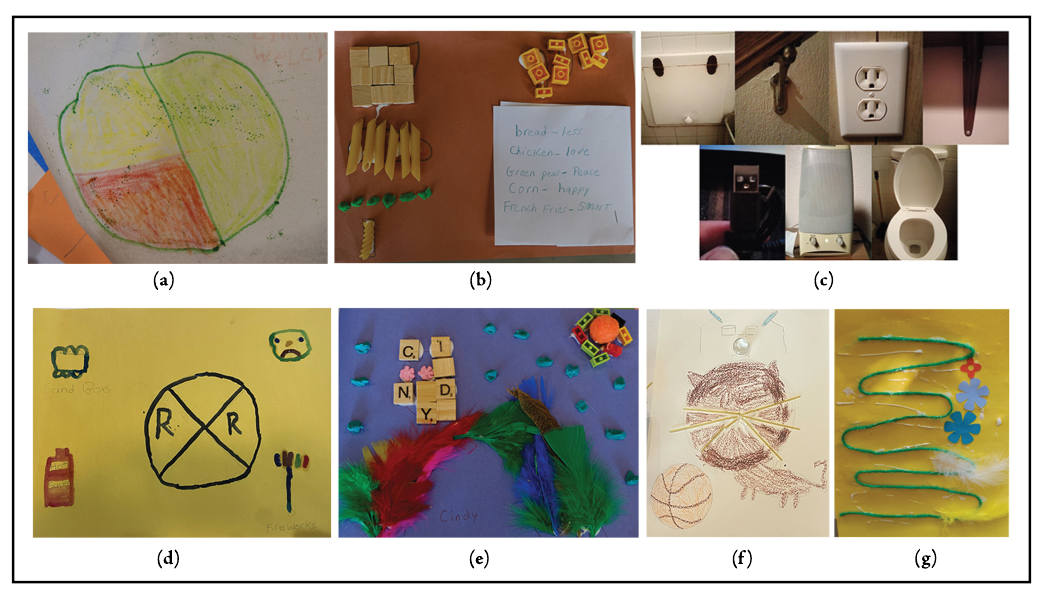}
\Description{A collage of artifacts made by participants, showing how they use visual analogies and metaphors to make data more understandable.}
\centering
\caption{Explaining data through visual analogies. Participants creatively transformed complex datasets into relatable and intuitive representations through imaginative and symbolic imagery. Employing metaphorical representation, they transformed the abstract concept of percentage into a tangible process akin to eating an apple, utilizing various colors to denote different stages (a). They associated materials with similar appearances, textures, or colors to create tactile representations using diverse materials (b, e). In addition, they symbolized abstract concepts, such as mood, by assigning facial expressions to objects, reflecting their mood for each day throughout the week (c). They also linked data with sensory experiences and their own identity, represented in their visualizations (d, f, g).} 
\label{fig:contexualize}
\end{figure*}

\section{Findings}
We collected over 240 artifacts from 20 participants across both workshops, along with video recordings and detailed observations from five facilitators. We developed a codebook to analyze these artifacts, focusing on materials used, physical characteristics, level of detail, and strategies or challenges in data authoring. We evaluated participants on their strengths, weaknesses, quality of work, completion time, effort, creativity, and identified their best and worst work. This analysis aimed to address our core research question: \textit{How do people with IDD approach and visualize data?} Below, we present general insights from the workshops, highlighting patterns in data sense-making and visualization.

\subsection{Exploring Data through Hands-on Activities}
Participants with IDD and their caregivers tended to employ a constructionist approach~\cite{fosnot1996constructivism} to data exploration and visualization. Instead of simple sketches, they built physical projects to represent data, engaging in physical tasks and using tangible items. They incorporated isotype visualization techniques~\cite{isotype} to make abstract data more concrete. This hands-on, constructive approach~\cite{huron2014constructive} facilitated direct interaction with data, enhancing understanding and engagement through physical manipulation.

In the Creative Selfie activity, one participant used various candies to map out their facial features, such as chocolate chips for eyes and gummy worms for the nose and lips (Figure \ref{fig:constructionist} (a)). By manipulating their favorite candies, the participant turned the abstract idea of identity into a fun, interactive learning experience. As they noted, ``I loved picking out the candies and seeing my face and personality come together.'' In the Happy List activity, another participant used pasta and beans to represent a recent family gathering (Figure \ref{fig:constructionist} (d)), stating, ``I will make a dish with things on my happy list.'' This example demonstrates how participants tend to think concretely, infusing data with meanings to represent personal stories and memories.

Using tangible materials also helped overcome communication challenges and fostered mutual understanding. In the Happy List activity (Figure \ref{fig:constructionist} (b)), a participant used colored stick figures to represent significant people in their life and pasta lines to show the strength of their bonds. This method highlighted the qualitative aspects of data, making its nuances and quality more visible and engaging. In the Pictorial Essay activity (Figure \ref{fig:constructionist} (c)), a caregiver helped a participant recall weekly events using tangible items: YMCA blocks for a trip, an orange pom-pom for a basketball game, a black button for bowling, green sticky notes for shopping, and a stick figure with a feather for a party.  These items made it easier for the participant to remember and share their experiences. As the caregiver noted, ``it's like each item brings back a specific memory.''

Throughout the workshop, participants frequently used isotype visualization~\cite{isotype}, which employs countable pictograms (Figure \ref{fig:constructionist}). This aligns with prior findings that isotypes are more accessible for people with IDD~\cite{chi21}, aiding working memory and linking abstract concepts to tangible quantities. Using various materials, isotypes integrated semantics and connected abstract data to countable quantities. For example, participants used this approach to express various preferences (Figure \ref{fig:constructionist} (e)), illustrate percentages (Figure \ref{fig:constructionist} (f)), and represent diet compositions (Figure \ref{fig:constructionist} (g)). This approach helped participants link numbers to meaning. As one participant said, ``Making the numbers concrete made it easier to understand.'' However, they also noted that this abstract representation did not come naturally: ``But I’d probably just have a picture of me doing that activity,'' highlighting the need for more intuitive and contextually rich data visualization.

These findings demonstrate the effectiveness of a \textit{constructionist} approach in data representation for people with IDD. Hands-on activities and physical objects helped participants give personal meaning to data, express themselves creatively, and make abstract concepts more concrete. Future research should explore several key directions. First, integrate constructionist approaches into data education curricula and encourage multisensory learning for people with IDD. Second, investigate how novel digital technologies can facilitate physical interactions with data to create more immersive and engaging experiences. Lastly, develop new design guidelines to explore data physicalization with various materials to improve comprehension and best practices.

\subsection{Humanizing Data with Personal Narratives}
Participants sought to personalize data by incorporating personally meaningful elements into their visualizations. This approach connected the data to their experiences, turning abstract information into relatable narratives and making visualization a medium for sharing personal stories and perspectives.

One strategy involved using symbols. For example, one participant frequently used the heart symbol (Figure \ref{fig:humanize} (a, b, c)). In the Creative Selfie, it represented love for family and friends, while in tasks like Circle Hunt and Circle Groups, it expressed frustration. Using personally meaningful symbols and imagery allowed participants to convey complex feelings in a way that resonated with them, fostering a deeper connection with the data being represented.

Participants also emphasized emotional elements in their data visualizations by linking data points to personal experiences and feelings. For example, in the Cats \& Dogs activity, participants used facts about these animals to create visualizations that taught key lessons. One participant, who had a traumatic experience with dogs, showed that dogs need much more care than cats, framing this as a negative (Figure \ref{fig:humanize} (f)). They commented, ``I had a bad experience with dogs and I always hated them, so I showed that in my visualization.'' This personal perspective added an emotional dimension to the data, illustrating how personal experiences can shape the design of visualization. 

Some participants took their visualizations a step further by developing characters that gave the data a human face and personality. This approach made the information more relatable and engaging by personifying abstract data points. For instance, in the Pictorial Essay task, one participant crafted a comic strip (Figure \ref{fig:humanize} (e)) detailing their week. In the Creative Selfie task, a participant drew an entire baseball field and team within a circle, using color to highlight player traits (Figure \ref{fig:humanize} (d)). Although unconventional for a selfie, this approach added personal meaning and depth to the data. As the participant explained, ``I love baseball, so I made it part of my project to show how it’s a big part of me.'' 

These observations demonstrate how participants with IDD used personal narratives to humanize data and convey their experiences through visualizations. By incorporating personalization, storytelling, and relating data to their everyday lives, they assigned personal values to numerical data, transforming abstract information into concrete human experiences. This underscores the role of \textit{affect} and \textit{storytelling} in improving data accessibility. To support accessible data authoring and creativity, future authoring tools could facilitate personalization by embedding features that allow users to add custom icons, symbols, photos, and personal elements, infusing visualizations with personal meaning and relevance. Additionally, visualization design guidelines should capture and represent the emotional context of data to enhance engagement. This involves expanding evaluation metrics to help users connect with the data on a deeper level and investigating what makes useful chartjunk~\cite{Bateman2010}. Finally, designers might incorporate characters, storylines, and leverage anthropomorphism~\cite{Lee2018} to personify data points. Collectively, these strategies aim to make data visualization more inclusive and empathetic, fostering a deeper connection between people and data, particularly benefiting individuals with cognitive disabilities.

\subsection{Explaining Data through Visual Analogies}

Participants faced significant challenges with numerical data, reflecting common IDD limitations~\cite{Lindstrm2023}. To overcome these, they used visual metaphors and analogies to simplify complex information, linking unfamiliar concepts with everyday experiences. This approach made abstract concepts more intuitive, allowing them to interpret and manipulate data more effectively.

One strategy involved using analogical reasoning, where participants translated conceptual mathematical problems into practical scenarios, echoing findings from research on health data literacy~\cite{hdv} and special education~\cite{math}. In the Color Challenge task, participants were asked to visualize three proportions. One participant depicted the process of eating an apple by drawing a circle and assigning colors—red for eaten, yellow for partially eaten, and green for untouched—to different stages (Figure \ref{fig:contexualize} (a)), creating something akin to a pie chart grounded in a real-world analog. Dividing the circle showed the percentages of the apple eaten, planned to be eaten, and remaining, linking abstract numbers to real-life.

Participants also used material metaphors to make data tangible and intuitive. In the Four Circle activity (Figure \ref{fig:contexualize} (e)), one participant created a scene with LEGO bricks for stability, an orange pom-pom for the Sun, colored feathers for a rainbow, and play-doh for raindrops. These materials, chosen for their distinct visual and tactile properties, created direct analogies to data qualities. The solidity of LEGO bricks symbolized stability, while the softness of the pom-pom and fluidity of play-doh represented variable or dynamic data. Bright colors provided clear visual differentiation, making data categories recognizable and engaging. Similarly, in the Happy List task (Figure \ref{fig:contexualize} (b)), a participant used wooden cubes for bread, LEGO bricks for chicken, penne for corn, play-doh for peas, and fusilli for fries. Words like “less,” “love,” “peace,” “happy,” and “smart” indicated different levels of happiness and feelings associated with each item, conveying the quality rather than quantity of the data, and adding extra layers of meaning to the visualization.

Another strategy involved using symbolic and sensory associations to create multisensory visualizations. In the Circle Hunt activity, a participant recalled circular objects that triggered intense sensory experiences, such as railroad crossings with loud noises, a frowning face for displeasure, hot sauces for a burning sensation, and fireworks for discomfort from bright lights and sounds. They represented these emotional responses as proxy data for the underlying categories reflected by the circles. 
In the Pictorial Essay activity, a participant selected seven photos from their life, each resembling a human face to convey their feelings for each day (Figure \ref{fig:contexualize} (c)). For example, Wednesday was shown with an electric outlet to signify an okay mood, Saturday with a speaker for excitement, and Sunday with a toilet for sadness about the upcoming week. This method shows how everyday objects can be used for creative data expression, animating abstract feelings and experiences.

These findings create novel design opportunities for data visualization, encouraging researchers and practitioners to explore innovative strategies for data encoding and communication. They highlight the need to move beyond statistical analysis and standardized abstract encodings to consider representations that offer analogs to how people experience data. Participants' analogous approach to data emphasizes intuitive representations that pair the inherent characteristics of the data with visual elements and design strategies that enhance semantic understanding. This means creating visualizations where attributes such as shape, color, texture, and spatial arrangement directly correspond to experiential features of the data, making thoughtful design decisions that convey deeper meaning and support visual reasoning.

Future work may continue this investigation in several ways. First, visualization design could enhance visual reasoning and semiotics by studying how users interpret visual analogies, metaphors, and symbols to create more effective and universally understandable visualizations. Second, researchers could develop novel representations and interactive formats beyond traditional charts, such as data comics~\cite{datacomics}, narrated data videos~\cite{Amini2015}, and gamified data experiences~\cite{gamedata}. These approaches provide engaging and intuitive ways to explore and consume data grounded more in experience than abstraction. Lastly, researchers could explore using multisensory elements (e.g., touch, sight, sound) in data representations to meet diverse learning and sensory needs. This could include haptic feedback devices, auditory cues, and multimedia elements to provide a richer sensory experience.

\section{Discussion} 
Visualization has transformative potential to empower individuals with IDD by making data more accessible~\cite{vis23, hdv}. However, we have limited guidance on designing cognitively accessible visualizations~\cite{chi21}, and such guidance is often grounded in assessments of conventional visualization approaches rather than the authentic expression of data created by people with IDD. This study aims to address this gap by engaging 20 individuals with IDD in a series of hybrid participatory design workshops to establish best practices for cognitive data accessibility.
Reflecting on participants’ strategies and challenges in approaching and visualizing data, we propose two future directions for cognitively accessible visualizations: transforming numbers into narratives and blending data design with everyday aesthetics. These approaches address critical aspects of accessible data representation, expression, interaction, and education. This section will explore these directions in detail, reimagining how visualization can enhance data accessibility and democratize data access for individuals with diverse cognitive abilities.

\subsection{Transforming Numbers into Narratives}
Consistent with past studies~\cite{math, chi21}, participants with IDD preferred concrete approaches to data reasoning and found data more understandable when directly related to their personal experiences. 
The abstract nature of data often caused frustration, highlighting the need for cognitively accessible and emotionally considerate data visualizations. Participants commonly deabstracted data by contextualizing and humanizing it with values, meanings, and narratives. Most of them struggled with grasping numerical data without concrete context. This suggests that narrative visualization~\cite{Hullman2011} may bridge the gap between abstract data and concrete understanding.

\subsubsection{Designing Relatable Data Narratives}
Storytelling was a preferred method for participants to consume data and create visualizations. Although storytelling with data is relatively well-established~\cite{segel2010narrative, Kosara2013}, we lack robust evidence on effective data story design for people with cognitive disabilities. Our workshops highlighted the importance of \textit{affect} and \textit{context} in storytelling. Emotional experiences shaped both the data participants chose to visualize and the way they represented it. Metaphors and analogies contextualized the data and conveyed personal insights and perspectives. These observations suggest that accessible data stories should be both emotionally responsive and contextually rich. Emotional responsiveness involves creating visualizations that connect with audiences on a deep, visceral level, while contextual richness embeds data in familiar, real-world scenarios. Future research could further explore affective design principles~\cite{LeeRobbins2022, Lan2024}, such as how emotional cues like color and imagery impact understanding and engagement. Additionally, narrative formats rich in metaphors and grounded in realistic contexts, such as comics~\cite{datacomics, personaldatacomics} and cinematics~\cite{Amini2015, Conlen2023CinematicTI}, might enhance data accessibility. Combining these methods may lead to more engaging and meaningful data narratives for individuals with IDD.

Participants often connected data to specific individuals by creating characters in their visualizations. These characters represented themselves, people and animals in their lives, or individuals they admired, situating abstract data in personally meaningful contexts. This highlights the potential of making data visualizations more personal and relatable through familiar characters. Future research could explore how integrating characters or human narrators in data visualizations enhances audience engagement and personal connection. Studies might investigate character-based narratives~\cite{character} for making data more approachable and examine the impact of anthropomorphic design~\cite{Lee2018} on emotional engagement. Such work could provide valuable insights into developing personalized and emotionally resonant data visualizations that better engage and support audiences.

Importantly, individuals with IDD should not be limited to high-level data summaries. Storytelling techniques can enhance visualizations by adding richer context and emotional resonance, improving understanding and engagement. We anticipate that narrative strategies will encourage deeper data engagement and provide effective scaffolding for better data comprehension and use.

\subsubsection{Supporting Personalized Data Expression}
Participants and caregivers frequently customized visualizations to meet diverse cognitive and sensory needs, highlighting the need for alternative visualizations and flexible authoring. Data was often represented in simplified, concrete formats, broken into smaller, manageable pieces. This preference underscores the benefits of modularity~\cite{Coltheart1999-nx} in accessible data authoring. Utilizing adjustable components—such as isotype visualizations~\cite{isotype} or LEGO-like building blocks~\cite{flack2019lego}—enables tailored visualizations, enhancing clarity and flexibility. Future research should explore how to utilize visual chunking strategies (i.e., grouping related data into distinct sections)~\cite{Stieff2020} and develop modular visualizations that can be adjusted and configured to better meet the needs of diverse audiences. Additionally, data authoring tools should offer customizable features and templates to encourage creativity and self-expression, allowing individuals to modify design elements, layouts, and interactive components to reflect their unique styles and preferences.

The study indicated that visualization could also serve as a tool for therapeutic expression, helping individuals process and communicate their emotions. For individuals with IDD, who often face challenges with verbal communication~\cite{criteria} and emotional regulation~\cite{emotionaldysregulation}, visualization provided a unique means to convey thoughts and feelings. Creating visualizations of personal data enabled non-verbal participants to express their emotions more effectively, uncover previously unrecognized insights about themselves, and build stronger connections with those around them. Notably, some initially reserved participants became more engaged as the workshop progressed, underscoring the therapeutic benefits of creative data expression. Future intelligent visualization tools could enhance personal data storytelling with features for guided reflection~\cite{Lee2017} and tracking emotional states over time~\cite{Nurain2023}. These advancements would provide personalized insights, helping individuals better understand their emotional journeys and express their experiences more clearly through their data stories.

Despite the benefits of customizable authoring tools and personalized analytics, such tools must be mindful of privacy and user agency. They should ensure robust data protection, give individuals with IDD easy access to and control over their data, and provide clear instructions to support their autonomy.

\subsection{Blending Data with Everyday Aesthetics}
Due to a historical lack of inclusion in traditional classrooms, individuals with IDD often have limited math and visualization literacy~\cite{chi21}. We observed that conventional visualizations, such as statistical charts, were unfamiliar to most of our participants. Instead of sticking to standardized chart types, participants and their caregivers experimented with diverse materials, sensory techniques, and visual formats in constructing data visualizations. In line with studies on inclusive design for learning and education~\cite{Polat2019, umnTIESTIPS, fosnot1996constructivism}, hands-on activities and constructionist learning~\cite{fosnot1996constructivism} helped participants understand abstract concepts significantly better. By manipulating objects, interacting with physical entities, and gradually building their understanding through activities like counting, sorting, and assembling, participants found abstract data became more concrete and easier to grasp. These approaches also reduced the intimidation associated with data~\cite{chi23}, making it more accessible and comfortable to work with. These observations suggest that physical data interactions~\cite{jansen2015opportunities, bae2022making} and multisensory engagement~\cite{lloyd2020multisensory} might enhance data accessibility for individuals with IDD. 

\subsubsection{Designing Organic Data Interaction}
Both online and in-person participants extensively utilized physicalization strategies in their visualizations. Online participants used household objects and photos, while in-person participants worked with materials like beans, beads, and play-doh, enabling hands-on interaction. Everyday objects fostered creative expression and helped with abstract reasoning. For example, in the Color Challenge activity, participants redesigned a pie chart to represent proportions. Although initially challenged by percentages, the metaphor of using plates proved to make such concepts more accessible. Dividing the plate into sections for different foods related the task to real life, and participants creatively used play-doh, colors, and various items to represent food groups and volumes. Feedback from in-person staff supports that familiar structures and starting points can enhance understanding for individuals with IDD. This highlights how linking abstract concepts to real-world experiences~\cite{chi21}, using concrete objects and templates, can make data more accessible and engaging.

Future visualization and accessibility research should further explore how everyday objects and real-world contexts can make data interactions more intuitive and relatable. For example, technologies like augmented reality~\cite{koushik2022towards} could blend data visualizations with physical spaces, making them more relevant. Tangible interfaces~\cite{Beccaluva2022} might allow individuals to engage directly with physical data representations, enhancing understanding. Modular and hybrid tools~\cite{Braga2011, bae2022cultivating} could combine digital and physical elements for personalized displays. Wearable technologies, such as smartwatches~\cite{While2024}, could offer real-time, glanceable data visualizations, while sensor-embedded items~\cite{GarcaPealvo2019} could provide dynamic, context-aware interactions. Exploring these approaches may lead to more accessible and effective data interactions.

Participants preferred visual and tangible reasoning over statistical methods when interpreting data, highlighting the benefits of integrating materiality and tangible metaphors into data representation. Physical objects with various textures, weights, and shapes provided sensory experiences that enhanced participants’ ability to engage with abstract concepts. These tactile elements made data more accessible and influenced how participants interacted with and understood it. Future work may further investigate how tangible data encoding methods can utilize the qualities of physical materials~\cite{Khot2014, khot2020shelfie} or human body movements~\cite{Payne2021} to design intuitive and accessible data interactions, combining tactile-kinesthetic interaction with cognitive reasoning.

\subsubsection{Supporting Multisensory Data Engagement}
Individuals with IDD often have complex needs~\cite{criteria}, heightened awareness of their surroundings, and are prone to emotional dysregulation~\cite{EngelYeger2011, Samson2014}. These characteristics require adapted strategies for data interaction, yet effective interventions remain limited. In our study, participants and caregivers employed multisensory techniques---visual, tactile, and auditory---during data exploration. This approach not only enhanced their understanding and retention of information but also enabled them to create visualizations that reflected their unique sensory experiences. In the Creative Selfie task, a participant used blueberries to form a frowning mouth and added headphones and sunglasses to indicate discomfort with bright lights and loud noises, expressing their sensitivity and ongoing unhappiness. In the Circle Hunt activity, a non-verbal participant initially struggled to locate circular objects. The caregiver used words and gestures to guide attention, drew circle examples to clarify the concept, and employed physical objects to show how to identify and handle the items. These examples highlight the crucial role of multisensory approaches in making data visualization and education more accessible. Tailored strategies can help individuals with IDD better understand and engage with complex information. Future research should build on advances in multisensory data interaction~\cite{chundury2021towards}, exploring how integrating visual, tactile, and auditory modalities can enhance accessible learning experiences. This may include developing learning tools with adaptable interfaces and flexible input methods~\cite{Jorge2001} to cater to diverse sensory preferences. Moreover, refining tangible learning strategies~\cite{Koushik2019} with physical and constructive elements can foster hands-on exploration. Together, these approaches could make data education more inclusive, addressing various sensory needs and fostering engagement.

Reflecting on both remote and in-person workshop formats, we recommend prioritizing in-person collaborative data learning experiences for students with IDD. While the online data pen-pal format maintained engagement through peer feedback, in-person small-group activities fostered better peer support and community building. Despite individual preferences and pacing, face-to-face interactions offer significant benefits for creating a stronger learning environment and enhancing collaboration.

\subsection{Limitations}
The limitations of this study primarily center on engaging individuals with IDD in understanding and working with data. Participants may have faced conceptual or emotional challenges due to the complexity of the data involved. Additionally, varying levels of data literacy among participants could have impacted their ability to engage effectively with the data and design visualizations. Communication difficulties may have made it challenging for participants to provide meaningful feedback on data visualizations or comment on the process. Moreover, we were unable to link specific participants to their disabilities, demographics, or other relevant factors, preventing us from gaining a granular understanding of how different types of IDD impact data authoring. Despite these challenges, our study successfully balanced adapting activities to participants’ needs with advancing our understanding of accessible data interactions, paving the way for future research in this field.

\section{Conclusion}
In this study, we involved 20 participants with IDD as co-designers to investigate their approaches to data visualization through a series of hybrid participatory design workshops. By analyzing the artifacts created by participants and reflecting on our observations, we identified three strategies they use to approach data and construct visualizations: hands-on exploration, personal narratives, and visual analogies. Based on these findings, we propose alternative guidance for cognitively accessible visualizations, emphasizing the transformation of complex data into compelling narratives and the integration of data design with everyday aesthetics.

\begin{acks}
We would like to thank the Enrichment Center for their participation. This work was supported by the Ray Hauser Award at the University of Colorado Boulder, NSF \#2320920, and DHHS Cooperative Agreement \#90DNPA00005-01-00.
\end{acks}

\bibliographystyle{ACM-Reference-Format}
\bibliography{idd}


\begin{thebibliography}{112}

%%% the Web address.

\ifx \showCODEN    \undefined \def \showCODEN     #1{\unskip}     \fi
\ifx \showDOI      \undefined \def \showDOI       #1{#1}\fi
\ifx \showISBNx    \undefined \def \showISBNx     #1{\unskip}     \fi
\ifx \showISBNxiii \undefined \def \showISBNxiii  #1{\unskip}     \fi
\ifx \showISSN     \undefined \def \showISSN      #1{\unskip}     \fi
\ifx \showLCCN     \undefined \def \showLCCN      #1{\unskip}     \fi
\ifx \shownote     \undefined \def \shownote      #1{#1}          \fi
\ifx \showarticletitle \undefined \def \showarticletitle #1{#1}   \fi
\ifx \showURL      \undefined \def \showURL       {\relax}        \fi
% invisible to TeX
\providecommand\bibfield[2]{#2}
\providecommand\bibinfo[2]{#2}
\providecommand\natexlab[1]{#1}
\providecommand\showeprint[2][]{arXiv:#2}

\bibitem[dea({[n.\,d.]})]%
        {deardata}
 \bibinfo{year}{[n.\,d.]}\natexlab{}.
\newblock \bibinfo{title}{Dear Data}.
\newblock
\newblock
\urldef\tempurl%
\url{https://www.dear-data.com}
\showURL{%
\tempurl}


\bibitem[cri({[n.\,d.]})]%
        {criteria}
 \bibinfo{year}{[n.\,d.]}\natexlab{}.
\newblock \bibinfo{title}{Defining Criteria for Intellectual Disability}.
\newblock
\newblock
\urldef\tempurl%
\url{https://www.aaidd.org/intellectual-disability/definition}
\showURL{%
\tempurl}


\bibitem[abo(2021)]%
        {about}
 \bibinfo{year}{2021}\natexlab{}.
\newblock \bibinfo{title}{About Intellectual and Developmental Disabilities}.
\newblock
\newblock
\urldef\tempurl%
\url{https://www.nichd.nih.gov/health/topics/idds/conditioninfo}
\showURL{%
\tempurl}


\bibitem[Acton et~al\mbox{.}(2023)]%
        {Acton2023}
\bibfield{author}{\bibinfo{person}{Daniel~James Acton}, \bibinfo{person}{Robert Waites}, \bibinfo{person}{Sujeet Jaydeokar}, {and} \bibinfo{person}{Steven Jones}.} \bibinfo{year}{2023}\natexlab{}.
\newblock \showarticletitle{Co-design and development of a multi-component anxiety management programme for people with an intellectual disability}.
\newblock \bibinfo{journal}{\emph{Advances in Mental Health and Intellectual Disabilities}} \bibinfo{volume}{17}, \bibinfo{number}{1} (\bibinfo{date}{Jan.} \bibinfo{year}{2023}), \bibinfo{pages}{26–36}.
\newblock
\showISSN{2044-1282}
\urldef\tempurl%
\url{https://doi.org/10.1108/amhid-04-2022-0017}
\showDOI{\tempurl}


\bibitem[Amini et~al\mbox{.}(2015)]%
        {Amini2015}
\bibfield{author}{\bibinfo{person}{Fereshteh Amini}, \bibinfo{person}{Nathalie Henry~Riche}, \bibinfo{person}{Bongshin Lee}, \bibinfo{person}{Christophe Hurter}, {and} \bibinfo{person}{Pourang Irani}.} \bibinfo{year}{2015}\natexlab{}.
\newblock \showarticletitle{Understanding Data Videos: Looking at Narrative Visualization through the Cinematography Lens}. In \bibinfo{booktitle}{\emph{Proceedings of the 33rd Annual ACM Conference on Human Factors in Computing Systems}} \emph{(\bibinfo{series}{CHI ’15})}. \bibinfo{publisher}{ACM}.
\newblock
\urldef\tempurl%
\url{https://doi.org/10.1145/2702123.2702431}
\showDOI{\tempurl}


\bibitem[Bach et~al\mbox{.}(2018)]%
        {datacomics}
\bibfield{author}{\bibinfo{person}{Benjamin Bach}, \bibinfo{person}{Zezhong Wang}, \bibinfo{person}{Matteo Farinella}, \bibinfo{person}{Dave Murray-Rust}, {and} \bibinfo{person}{Nathalie Henry~Riche}.} \bibinfo{year}{2018}\natexlab{}.
\newblock \showarticletitle{Design Patterns for Data Comics}. \bibinfo{pages}{1--12}.
\newblock
\urldef\tempurl%
\url{https://doi.org/10.1145/3173574.3173612}
\showDOI{\tempurl}


\bibitem[Bae et~al\mbox{.}(2022a)]%
        {bae2022cultivating}
\bibfield{author}{\bibinfo{person}{S~Sandra Bae}, \bibinfo{person}{Rishi Vanukuru}, \bibinfo{person}{Ruhan Yang}, \bibinfo{person}{Peter Gyory}, \bibinfo{person}{Ran Zhou}, \bibinfo{person}{Ellen Yi-Luen Do}, {and} \bibinfo{person}{Danielle~Albers Szafir}.} \bibinfo{year}{2022}\natexlab{a}.
\newblock \showarticletitle{Cultivating visualization literacy for children through curiosity and play}.
\newblock \bibinfo{journal}{\emph{IEEE Transactions on Visualization and Computer Graphics}} \bibinfo{volume}{29}, \bibinfo{number}{1} (\bibinfo{year}{2022}), \bibinfo{pages}{257--267}.
\newblock


\bibitem[Bae et~al\mbox{.}(2022b)]%
        {bae2022making}
\bibfield{author}{\bibinfo{person}{S~Sandra Bae}, \bibinfo{person}{Clement Zheng}, \bibinfo{person}{Mary~Etta West}, \bibinfo{person}{Ellen Yi-Luen Do}, \bibinfo{person}{Samuel Huron}, {and} \bibinfo{person}{Danielle~Albers Szafir}.} \bibinfo{year}{2022}\natexlab{b}.
\newblock \showarticletitle{Making data tangible: A cross-disciplinary design space for data physicalization}. In \bibinfo{booktitle}{\emph{Proceedings of the 2022 CHI Conference on Human Factors in Computing Systems}}. \bibinfo{pages}{1--18}.
\newblock


\bibitem[Bateman et~al\mbox{.}(2010)]%
        {Bateman2010}
\bibfield{author}{\bibinfo{person}{Scott Bateman}, \bibinfo{person}{Regan~L. Mandryk}, \bibinfo{person}{Carl Gutwin}, \bibinfo{person}{Aaron Genest}, \bibinfo{person}{David McDine}, {and} \bibinfo{person}{Christopher Brooks}.} \bibinfo{year}{2010}\natexlab{}.
\newblock \showarticletitle{Useful junk?: the effects of visual embellishment on comprehension and memorability of charts}. In \bibinfo{booktitle}{\emph{Proceedings of the SIGCHI Conference on Human Factors in Computing Systems}} \emph{(\bibinfo{series}{CHI ’10})}. \bibinfo{publisher}{ACM}.
\newblock
\urldef\tempurl%
\url{https://doi.org/10.1145/1753326.1753716}
\showDOI{\tempurl}


\bibitem[Bayor(2019)]%
        {bayor2019howtoapp}
\bibfield{author}{\bibinfo{person}{Andrew~A Bayor}.} \bibinfo{year}{2019}\natexlab{}.
\newblock \showarticletitle{HowToApp: Supporting life skills development of young adults with intellectual disability}. In \bibinfo{booktitle}{\emph{Proceedings of the 21st International ACM SIGACCESS Conference on Computers and Accessibility}}. \bibinfo{pages}{697--699}.
\newblock


\bibitem[Beccaluva et~al\mbox{.}(2022)]%
        {Beccaluva2022}
\bibfield{author}{\bibinfo{person}{Eleonora Beccaluva}, \bibinfo{person}{Fabiano Riccardi}, \bibinfo{person}{Mattia Gianotti}, \bibinfo{person}{Jessica Barbieri}, {and} \bibinfo{person}{Franca Garzotto}.} \bibinfo{year}{2022}\natexlab{}.
\newblock \showarticletitle{VIC — A Tangible User Interface to train memory skills in children with Intellectual Disability}.
\newblock \bibinfo{journal}{\emph{International Journal of Child-Computer Interaction}}  \bibinfo{volume}{32} (\bibinfo{date}{June} \bibinfo{year}{2022}), \bibinfo{pages}{100376}.
\newblock
\showISSN{2212-8689}
\urldef\tempurl%
\url{https://doi.org/10.1016/j.ijcci.2021.100376}
\showDOI{\tempurl}


\bibitem[Beheshti et~al\mbox{.}(2020)]%
        {iStory}
\bibfield{author}{\bibinfo{person}{Amin Beheshti}, \bibinfo{person}{Alireza Tabebordbar}, {and} \bibinfo{person}{Boualem Benatallah}.} \bibinfo{year}{2020}\natexlab{}.
\newblock \showarticletitle{iStory: Intelligent Storytelling with Social Data}. In \bibinfo{booktitle}{\emph{Companion Proceedings of the Web Conference 2020}} (Taipei, Taiwan) \emph{(\bibinfo{series}{WWW '20})}. \bibinfo{publisher}{Association for Computing Machinery}, \bibinfo{address}{New York, NY, USA}, \bibinfo{pages}{253–256}.
\newblock
\showISBNx{9781450370240}
\urldef\tempurl%
\url{https://doi.org/10.1145/3366424.3383553}
\showDOI{\tempurl}


\bibitem[Bietti et~al\mbox{.}(2018)]%
        {collective}
\bibfield{author}{\bibinfo{person}{Lucas Bietti}, \bibinfo{person}{Ottilie Tilston}, {and} \bibinfo{person}{Adrian Bangerter}.} \bibinfo{year}{2018}\natexlab{}.
\newblock \showarticletitle{Storytelling as Adaptive Collective Sensemaking}.
\newblock \bibinfo{journal}{\emph{Topics in Cognitive Science}}  \bibinfo{volume}{11} (\bibinfo{date}{06} \bibinfo{year}{2018}).
\newblock
\urldef\tempurl%
\url{https://doi.org/10.1111/tops.12358}
\showDOI{\tempurl}


\bibitem[Boy et~al\mbox{.}(2015)]%
        {engagement}
\bibfield{author}{\bibinfo{person}{Jeremy Boy}, \bibinfo{person}{Francoise Detienne}, {and} \bibinfo{person}{Jean-Daniel Fekete}.} \bibinfo{year}{2015}\natexlab{}.
\newblock \showarticletitle{Storytelling in Information Visualizations: Does it Engage Users to Explore Data?}. In \bibinfo{booktitle}{\emph{Proceedings of the 33rd Annual ACM Conference on Human Factors in Computing Systems}} (Seoul, Republic of Korea) \emph{(\bibinfo{series}{CHI '15})}. \bibinfo{publisher}{Association for Computing Machinery}, \bibinfo{address}{New York, NY, USA}, \bibinfo{pages}{1449–1458}.
\newblock
\showISBNx{9781450331456}
\urldef\tempurl%
\url{https://doi.org/10.1145/2702123.2702452}
\showDOI{\tempurl}


\bibitem[Braga et~al\mbox{.}(2011)]%
        {Braga2011}
\bibfield{author}{\bibinfo{person}{Rodrigo Antonio~Marques Braga}, \bibinfo{person}{Marcelo Petry}, \bibinfo{person}{Luis~Paulo Reis}, {and} \bibinfo{person}{Antnio~Paulo Moreira}.} \bibinfo{year}{2011}\natexlab{}.
\newblock \showarticletitle{IntellWheels: Modular development platform for intelligent wheelchairs}.
\newblock \bibinfo{journal}{\emph{The Journal of Rehabilitation Research and Development}} \bibinfo{volume}{48}, \bibinfo{number}{9} (\bibinfo{year}{2011}), \bibinfo{pages}{1061}.
\newblock
\showISSN{0748-7711}
\urldef\tempurl%
\url{https://doi.org/10.1682/jrrd.2010.08.0139}
\showDOI{\tempurl}


\bibitem[Bratitsis(2016)]%
        {digital}
\bibfield{author}{\bibinfo{person}{Tharrenos Bratitsis}.} \bibinfo{year}{2016}\natexlab{}.
\newblock \showarticletitle{A Digital Storytelling Approach for Fostering Empathy Towards Autistic Children: Lessons learned} \emph{(\bibinfo{series}{DSAI '16})}. \bibinfo{publisher}{Association for Computing Machinery}, \bibinfo{address}{New York, NY, USA}, \bibinfo{pages}{301–308}.
\newblock
\showISBNx{9781450347488}
\urldef\tempurl%
\url{https://doi.org/10.1145/3019943.3019987}
\showDOI{\tempurl}


\bibitem[Bratitsis and Ziannas(2015)]%
        {socialempathy}
\bibfield{author}{\bibinfo{person}{Tharrenos Bratitsis} {and} \bibinfo{person}{Petros Ziannas}.} \bibinfo{year}{2015}\natexlab{}.
\newblock \showarticletitle{From Early Childhood to Special Education: Interactive Digital Storytelling as a Coaching Approach for Fostering Social Empathy}.
\newblock \bibinfo{journal}{\emph{Procedia Computer Science}}  \bibinfo{volume}{67} (\bibinfo{year}{2015}), \bibinfo{pages}{231--240}.
\newblock
\showISSN{1877-0509}
\urldef\tempurl%
\url{https://doi.org/10.1016/j.procs.2015.09.267}
\showDOI{\tempurl}
\newblock
\shownote{Proceedings of the 6th International Conference on Software Development and Technologies for Enhancing Accessibility and Fighting Info-exclusion}.


\bibitem[Braun and Clarke(2012)]%
        {braun2012thematic}
\bibfield{author}{\bibinfo{person}{Virginia Braun} {and} \bibinfo{person}{Victoria Clarke}.} \bibinfo{year}{2012}\natexlab{}.
\newblock \bibinfo{booktitle}{\emph{Thematic analysis.}}
\newblock \bibinfo{publisher}{American Psychological Association}.
\newblock


\bibitem[Brooks(1987)]%
        {therapeutic}
\bibfield{author}{\bibinfo{person}{Robert~B. Brooks}.} \bibinfo{year}{1987}\natexlab{}.
\newblock \showarticletitle{Storytelling and the Therapeutic Process for Children with Learning Disabilities}.
\newblock \bibinfo{journal}{\emph{Journal of Learning Disabilities}} \bibinfo{volume}{20}, \bibinfo{number}{9} (\bibinfo{year}{1987}), \bibinfo{pages}{546--550}.
\newblock
\urldef\tempurl%
\url{https://doi.org/10.1177/002221948702000906}
\showDOI{\tempurl}
\showeprint{https://doi.org/10.1177/002221948702000906}
\newblock
\shownote{PMID: 3694046}.


\bibitem[Brug et~al\mbox{.}(2013)]%
        {learnandapply}
\bibfield{author}{\bibinfo{person}{Annet Brug}, \bibinfo{person}{Annette Putten}, {and} \bibinfo{person}{Carla Vlaskamp}.} \bibinfo{year}{2013}\natexlab{}.
\newblock \showarticletitle{Learn and apply: Using multi-sensory storytelling to gather knowledge about preferences and abilities of children with profound intellectual and multiple disabilities - Three case studies}.
\newblock \bibinfo{journal}{\emph{Journal of intellectual disabilities : JOID}} (\bibinfo{date}{10} \bibinfo{year}{2013}).
\newblock
\urldef\tempurl%
\url{https://doi.org/10.1177/1744629513508384}
\showDOI{\tempurl}


\bibitem[Card et~al\mbox{.}(1999)]%
        {vision_think}
\bibfield{author}{\bibinfo{person}{Stuart Card}, \bibinfo{person}{Jock Mackinlay}, {and} \bibinfo{person}{Ben Shneiderman}.} \bibinfo{year}{1999}\natexlab{}.
\newblock \bibinfo{booktitle}{\emph{Readings in Information Visualization: Using Vision To Think}}.
\newblock \bibinfo{publisher}{Academic Press}.
\newblock
\showISBNx{978-1-55860-533-6}


\bibitem[Center(2024)]%
        {umnTIESTIPS}
\bibfield{author}{\bibinfo{person}{UMN~TIES Center}.} \bibinfo{year}{2024}\natexlab{}.
\newblock \bibinfo{title}{The Use of Graphic Organizers in Inclusive Classrooms for Students with Significant Cognitive Disabilities}.
\newblock \bibinfo{howpublished}{\url{https://publications.ici.umn.edu/ties/foundations-of-inclusion-tips/graphic-organizers-in-inclusive-classrooms-for-students-with-significant-cognitive-disabilities}}.
\newblock
\newblock
\shownote{[Accessed 23-07-2024]}.


\bibitem[Chundury et~al\mbox{.}(2021)]%
        {chundury2021towards}
\bibfield{author}{\bibinfo{person}{Pramod Chundury}, \bibinfo{person}{Biswaksen Patnaik}, \bibinfo{person}{Yasmin Reyazuddin}, \bibinfo{person}{Christine Tang}, \bibinfo{person}{Jonathan Lazar}, {and} \bibinfo{person}{Niklas Elmqvist}.} \bibinfo{year}{2021}\natexlab{}.
\newblock \showarticletitle{Towards understanding sensory substitution for accessible visualization: An interview study}.
\newblock \bibinfo{journal}{\emph{IEEE transactions on visualization and computer graphics}} \bibinfo{volume}{28}, \bibinfo{number}{1} (\bibinfo{year}{2021}), \bibinfo{pages}{1084--1094}.
\newblock


\bibitem[Colin~Gibson et~al\mbox{.}(2020)]%
        {colin2020lessons}
\bibfield{author}{\bibinfo{person}{Ryan Colin~Gibson}, \bibinfo{person}{Mark D.~Dunlop}, {and} \bibinfo{person}{Matt-Mouley Bouamrane}.} \bibinfo{year}{2020}\natexlab{}.
\newblock \showarticletitle{Lessons from expert focus groups on how to better support adults with mild intellectual disabilities to engage in co-design}. In \bibinfo{booktitle}{\emph{Proceedings of the 22nd International ACM SIGACCESS Conference on Computers and Accessibility}}. \bibinfo{pages}{1--12}.
\newblock


\bibitem[Coltheart(1999)]%
        {Coltheart1999-nx}
\bibfield{author}{\bibinfo{person}{M Coltheart}.} \bibinfo{year}{1999}\natexlab{}.
\newblock \showarticletitle{Modularity and cognition}.
\newblock \bibinfo{journal}{\emph{Trends Cogn. Sci.}} \bibinfo{volume}{3}, \bibinfo{number}{3} (\bibinfo{date}{March} \bibinfo{year}{1999}), \bibinfo{pages}{115--120}.
\newblock


\bibitem[Conlen et~al\mbox{.}(2023)]%
        {Conlen2023CinematicTI}
\bibfield{author}{\bibinfo{person}{Matthew Conlen}, \bibinfo{person}{Jeffrey Heer}, \bibinfo{person}{Hillary Mushkin}, {and} \bibinfo{person}{Scott Davidoff}.} \bibinfo{year}{2023}\natexlab{}.
\newblock \showarticletitle{Cinematic Techniques in Narrative Visualization}.
\newblock \bibinfo{journal}{\emph{ArXiv}}  \bibinfo{volume}{abs/2301.03109} (\bibinfo{year}{2023}).
\newblock
\urldef\tempurl%
\url{https://api.semanticscholar.org/CorpusID:255546525}
\showURL{%
\tempurl}


\bibitem[Dasu et~al\mbox{.}(2024)]%
        {character}
\bibfield{author}{\bibinfo{person}{Keshav Dasu}, \bibinfo{person}{Yun-Hsin Kuo}, {and} \bibinfo{person}{Kwan-Liu Ma}.} \bibinfo{year}{2024}\natexlab{}.
\newblock \showarticletitle{Character-Oriented Design for Visual Data Storytelling}.
\newblock \bibinfo{journal}{\emph{IEEE Transactions on Visualization and Computer Graphics}} \bibinfo{volume}{30}, \bibinfo{number}{1} (\bibinfo{date}{Jan} \bibinfo{year}{2024}), \bibinfo{pages}{98--108}.
\newblock
\showISSN{1941-0506}


\bibitem[Engel-Yeger et~al\mbox{.}(2011)]%
        {EngelYeger2011}
\bibfield{author}{\bibinfo{person}{Batya Engel-Yeger}, \bibinfo{person}{Reem Hardal-Nasser}, {and} \bibinfo{person}{Eynat Gal}.} \bibinfo{year}{2011}\natexlab{}.
\newblock \showarticletitle{Sensory processing dysfunctions as expressed among children with different severities of intellectual developmental disabilities}.
\newblock \bibinfo{journal}{\emph{Research in Developmental Disabilities}} \bibinfo{volume}{32}, \bibinfo{number}{5} (\bibinfo{date}{Sept.} \bibinfo{year}{2011}), \bibinfo{pages}{1770–1775}.
\newblock
\showISSN{0891-4222}
\urldef\tempurl%
\url{https://doi.org/10.1016/j.ridd.2011.03.005}
\showDOI{\tempurl}


\bibitem[Erete et~al\mbox{.}(2016)]%
        {npo}
\bibfield{author}{\bibinfo{person}{Sheena Erete}, \bibinfo{person}{Emily Ryou}, \bibinfo{person}{Geoff Smith}, \bibinfo{person}{Khristina~Marie Fassett}, {and} \bibinfo{person}{Sarah Duda}.} \bibinfo{year}{2016}\natexlab{}.
\newblock \showarticletitle{Storytelling with Data: Examining the Use of Data by Non-Profit Organizations}. In \bibinfo{booktitle}{\emph{Proceedings of the 19th ACM Conference on Computer-Supported Cooperative Work \& Social Computing}} (San Francisco, California, USA) \emph{(\bibinfo{series}{CSCW '16})}. \bibinfo{publisher}{Association for Computing Machinery}, \bibinfo{address}{New York, NY, USA}, \bibinfo{pages}{1273–1283}.
\newblock
\showISBNx{9781450335928}
\urldef\tempurl%
\url{https://doi.org/10.1145/2818048.2820068}
\showDOI{\tempurl}


\bibitem[Flack et~al\mbox{.}(2019)]%
        {flack2019lego}
\bibfield{author}{\bibinfo{person}{Stuart Flack}, \bibinfo{person}{Kevin Ponto}, \bibinfo{person}{Travis Tangen}, {and} \bibinfo{person}{Karen~B Schloss}.} \bibinfo{year}{2019}\natexlab{}.
\newblock \showarticletitle{LEGO as Language for Visual Communication}.
\newblock  (\bibinfo{year}{2019}).
\newblock


\bibitem[Fosnot and Perry(1996)]%
        {fosnot1996constructivism}
\bibfield{author}{\bibinfo{person}{Catherine~Twomey Fosnot} {and} \bibinfo{person}{Randall~Stewart Perry}.} \bibinfo{year}{1996}\natexlab{}.
\newblock \showarticletitle{Constructivism: A psychological theory of learning}.
\newblock \bibinfo{journal}{\emph{Constructivism: Theory, perspectives, and practice}} \bibinfo{volume}{2}, \bibinfo{number}{1} (\bibinfo{year}{1996}), \bibinfo{pages}{8--33}.
\newblock


\bibitem[Frid et~al\mbox{.}(2022)]%
        {proxy}
\bibfield{author}{\bibinfo{person}{Emma Frid}, \bibinfo{person}{Claudio Panariello}, {and} \bibinfo{person}{Claudia Núñez-Pacheco}.} \bibinfo{year}{2022}\natexlab{}.
\newblock \showarticletitle{Customizing and Evaluating Accessible Multisensory Music Experiences with Pre-Verbal Children—A Case Study on the Perception of Musical Haptics Using Participatory Design with Proxies}.
\newblock \bibinfo{journal}{\emph{Multimodal Technologies and Interaction}} \bibinfo{volume}{6}, \bibinfo{number}{7} (\bibinfo{year}{2022}).
\newblock
\showISSN{2414-4088}
\urldef\tempurl%
\url{https://doi.org/10.3390/mti6070055}
\showDOI{\tempurl}


\bibitem[García-Peñalvo and Franco-Martín(2019)]%
        {GarcaPealvo2019}
\bibfield{author}{\bibinfo{person}{Francisco~José García-Peñalvo} {and} \bibinfo{person}{Manuel Franco-Martín}.} \bibinfo{year}{2019}\natexlab{}.
\newblock \showarticletitle{Sensor Technologies for Caring People with Disabilities}.
\newblock \bibinfo{journal}{\emph{Sensors}} \bibinfo{volume}{19}, \bibinfo{number}{22} (\bibinfo{date}{Nov.} \bibinfo{year}{2019}), \bibinfo{pages}{4914}.
\newblock
\showISSN{1424-8220}
\urldef\tempurl%
\url{https://doi.org/10.3390/s19224914}
\showDOI{\tempurl}


\bibitem[Glaser and Strauss(1967)]%
        {glaser1967discovery}
\bibfield{author}{\bibinfo{person}{B.G. Glaser} {and} \bibinfo{person}{A.L. Strauss}.} \bibinfo{year}{1967}\natexlab{}.
\newblock \bibinfo{booktitle}{\emph{The Discovery of Grounded Theory: Strategies for Qualitative Research}}.
\newblock \bibinfo{publisher}{Aldine}.
\newblock
\showISBNx{9780202302607}
\showLCCN{76434717}
\urldef\tempurl%
\url{https://books.google.com/books?id=oUxEAQAAIAAJ}
\showURL{%
\tempurl}


\bibitem[G\'{o}mez~Ortega et~al\mbox{.}(2024)]%
        {personaldatacomics}
\bibfield{author}{\bibinfo{person}{Alejandra G\'{o}mez~Ortega}, \bibinfo{person}{Jacky Bourgeois}, {and} \bibinfo{person}{Gerd Kortuem}.} \bibinfo{year}{2024}\natexlab{}.
\newblock \showarticletitle{Personal Data Comics: A Data Storytelling Approach Supporting Personal Data Literacy}. In \bibinfo{booktitle}{\emph{Proceedings of the XI Latin American Conference on Human Computer Interaction}} (<conf-loc>, <city>Puebla</city>, <country>Mexico</country>, </conf-loc>) \emph{(\bibinfo{series}{CLIHC '23})}. \bibinfo{publisher}{Association for Computing Machinery}, \bibinfo{address}{New York, NY, USA}, Article \bibinfo{articleno}{2}, \bibinfo{numpages}{8}~pages.
\newblock
\showISBNx{9798400716577}
\urldef\tempurl%
\url{https://doi.org/10.1145/3630970.3630982}
\showDOI{\tempurl}


\bibitem[Haroz et~al\mbox{.}(2015)]%
        {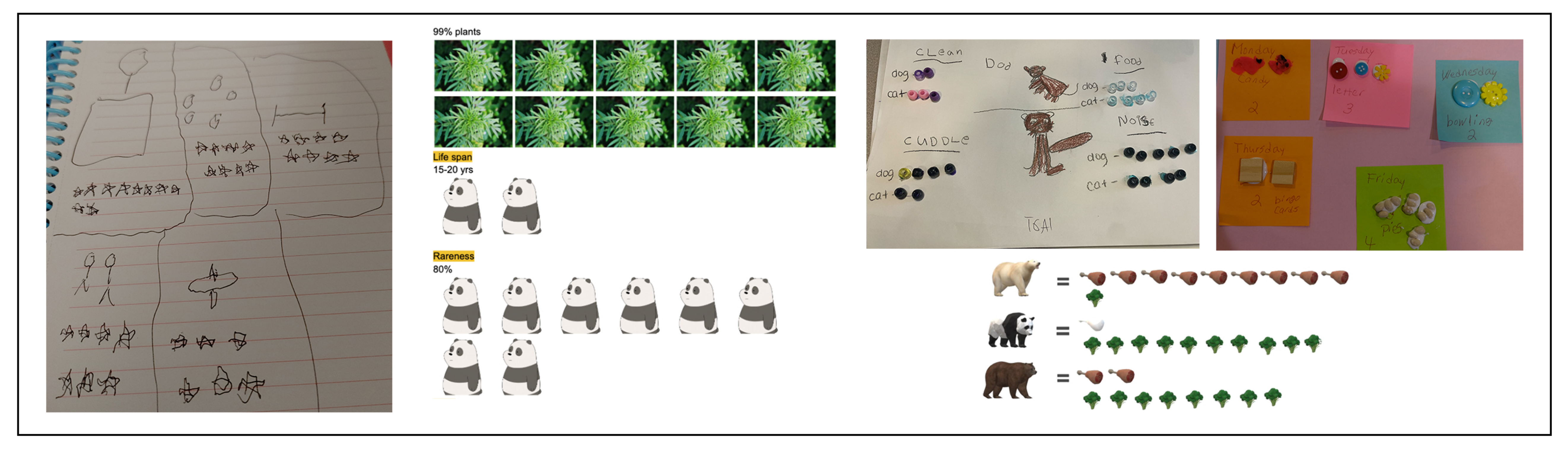}
\bibfield{author}{\bibinfo{person}{Steve Haroz}, \bibinfo{person}{Robert Kosara}, {and} \bibinfo{person}{Steven~L. Franconeri}.} \bibinfo{year}{2015}\natexlab{}.
\newblock \showarticletitle{ISOTYPE Visualization: Working Memory, Performance, and Engagement with Pictographs}. In \bibinfo{booktitle}{\emph{Proceedings of the 33rd Annual ACM Conference on Human Factors in Computing Systems}} (Seoul, Republic of Korea) \emph{(\bibinfo{series}{CHI '15})}. \bibinfo{publisher}{Association for Computing Machinery}, \bibinfo{address}{New York, NY, USA}, \bibinfo{pages}{1191–1200}.
\newblock
\showISBNx{9781450331456}
\urldef\tempurl%
\url{https://doi.org/10.1145/2702123.2702275}
\showDOI{\tempurl}


\bibitem[Heerings et~al\mbox{.}(2022)]%
        {heerings2022ask}
\bibfield{author}{\bibinfo{person}{Marjolijn Heerings}, \bibinfo{person}{Hester van~de Bovenkamp}, \bibinfo{person}{Mieke Cardol}, {and} \bibinfo{person}{Roland Bal}.} \bibinfo{year}{2022}\natexlab{}.
\newblock \showarticletitle{Ask us! Adjusting experience-based codesign to be responsive to people with intellectual disabilities, serious mental illness or older persons receiving support with independent living}.
\newblock \bibinfo{journal}{\emph{Health Expectations}} \bibinfo{volume}{25}, \bibinfo{number}{5} (\bibinfo{year}{2022}), \bibinfo{pages}{2246--2254}.
\newblock


\bibitem[Hendriks et~al\mbox{.}(2015)]%
        {hendriks2015codesign}
\bibfield{author}{\bibinfo{person}{Niels Hendriks}, \bibinfo{person}{Karin Slegers}, {and} \bibinfo{person}{Pieter Duysburgh}.} \bibinfo{year}{2015}\natexlab{}.
\newblock \showarticletitle{Codesign with people living with cognitive or sensory impairments: a case for method stories and uniqueness}.
\newblock \bibinfo{journal}{\emph{CoDesign}} \bibinfo{volume}{11}, \bibinfo{number}{1} (\bibinfo{year}{2015}), \bibinfo{pages}{70--82}.
\newblock


\bibitem[How et~al\mbox{.}(2017)]%
        {how2017envisioning}
\bibfield{author}{\bibinfo{person}{Tuck-Voon How}, \bibinfo{person}{Amy~S Hwang}, \bibinfo{person}{Robin~EA Green}, {and} \bibinfo{person}{Alex Mihailidis}.} \bibinfo{year}{2017}\natexlab{}.
\newblock \showarticletitle{Envisioning future cognitive telerehabilitation technologies: a co-design process with clinicians}.
\newblock \bibinfo{journal}{\emph{Disability and Rehabilitation: Assistive Technology}} \bibinfo{volume}{12}, \bibinfo{number}{3} (\bibinfo{year}{2017}), \bibinfo{pages}{244--261}.
\newblock


\bibitem[Hullman and Diakopoulos(2011)]%
        {Hullman2011}
\bibfield{author}{\bibinfo{person}{J. Hullman} {and} \bibinfo{person}{N. Diakopoulos}.} \bibinfo{year}{2011}\natexlab{}.
\newblock \showarticletitle{Visualization Rhetoric: Framing Effects in Narrative Visualization}.
\newblock \bibinfo{journal}{\emph{IEEE Transactions on Visualization and Computer Graphics}} \bibinfo{volume}{17}, \bibinfo{number}{12} (\bibinfo{date}{Dec.} \bibinfo{year}{2011}), \bibinfo{pages}{2231–2240}.
\newblock
\showISSN{1077-2626}
\urldef\tempurl%
\url{https://doi.org/10.1109/tvcg.2011.255}
\showDOI{\tempurl}


\bibitem[Huron et~al\mbox{.}(2014a)]%
        {constructive}
\bibfield{author}{\bibinfo{person}{Samuel Huron}, \bibinfo{person}{Sheelagh Carpendale}, \bibinfo{person}{Alice Thudt}, \bibinfo{person}{Anthony Tang}, {and} \bibinfo{person}{Michael Mauerer}.} \bibinfo{year}{2014}\natexlab{a}.
\newblock \showarticletitle{Constructive Visualization}.
\newblock \bibinfo{journal}{\emph{Proceedings of the Conference on Designing Interactive Systems: Processes, Practices, Methods, and Techniques, DIS}} (\bibinfo{date}{06} \bibinfo{year}{2014}).
\newblock
\showISBNx{978-1-4503-2902-6}
\urldef\tempurl%
\url{https://doi.org/10.1145/2598510.2598566}
\showDOI{\tempurl}


\bibitem[Huron et~al\mbox{.}(2014b)]%
        {huron2014constructive}
\bibfield{author}{\bibinfo{person}{Samuel Huron}, \bibinfo{person}{Sheelagh Carpendale}, \bibinfo{person}{Alice Thudt}, \bibinfo{person}{Anthony Tang}, {and} \bibinfo{person}{Michael Mauerer}.} \bibinfo{year}{2014}\natexlab{b}.
\newblock \showarticletitle{Constructive visualization}. In \bibinfo{booktitle}{\emph{Proceedings of the 2014 conference on Designing interactive systems}}. \bibinfo{pages}{433--442}.
\newblock


\bibitem[Huron et~al\mbox{.}(2014c)]%
        {constructive_token}
\bibfield{author}{\bibinfo{person}{Samuel Huron}, \bibinfo{person}{Yvonne Jansen}, {and} \bibinfo{person}{Sheelagh Carpendale}.} \bibinfo{year}{2014}\natexlab{c}.
\newblock \showarticletitle{Constructing Visual Representations: Investigating the Use of Tangible Tokens}.
\newblock \bibinfo{journal}{\emph{IEEE Transactions on Visualization and Computer Graphics}} \bibinfo{volume}{20}, \bibinfo{number}{12} (\bibinfo{date}{Dec} \bibinfo{year}{2014}), \bibinfo{pages}{2102--2111}.
\newblock
\showISSN{1941-0506}
\urldef\tempurl%
\url{https://doi.org/10.1109/TVCG.2014.2346292}
\showDOI{\tempurl}


\bibitem[Jagtap(2022)]%
        {marginalized}
\bibfield{author}{\bibinfo{person}{Santosh Jagtap}.} \bibinfo{year}{2022}\natexlab{}.
\newblock \showarticletitle{Co-design with marginalised people: designers’ perceptions of barriers and enablers}.
\newblock \bibinfo{journal}{\emph{CoDesign}} \bibinfo{volume}{18}, \bibinfo{number}{3} (\bibinfo{year}{2022}), \bibinfo{pages}{279--302}.
\newblock
\urldef\tempurl%
\url{https://doi.org/10.1080/15710882.2021.1883065}
\showDOI{\tempurl}
\showeprint{https://doi.org/10.1080/15710882.2021.1883065}


\bibitem[Jansen et~al\mbox{.}(2015)]%
        {jansen2015opportunities}
\bibfield{author}{\bibinfo{person}{Yvonne Jansen}, \bibinfo{person}{Pierre Dragicevic}, \bibinfo{person}{Petra Isenberg}, \bibinfo{person}{Jason Alexander}, \bibinfo{person}{Abhijit Karnik}, \bibinfo{person}{Johan Kildal}, \bibinfo{person}{Sriram Subramanian}, {and} \bibinfo{person}{Kasper Hornb{\ae}k}.} \bibinfo{year}{2015}\natexlab{}.
\newblock \showarticletitle{Opportunities and challenges for data physicalization}. In \bibinfo{booktitle}{\emph{proceedings of the 33rd annual acm conference on human factors in computing systems}}. \bibinfo{pages}{3227--3236}.
\newblock


\bibitem[Jorge(2001)]%
        {Jorge2001}
\bibfield{author}{\bibinfo{person}{Joaquim~A Jorge}.} \bibinfo{year}{2001}\natexlab{}.
\newblock \showarticletitle{Adaptive tools for the elderly: new devices to cope with age-induced cognitive disabilities}. In \bibinfo{booktitle}{\emph{Proceedings of the 2001 EC/NSF workshop on Universal accessibility of ubiquitous computing: providing for the elderly}} \emph{(\bibinfo{series}{WUAUC01})}. \bibinfo{publisher}{ACM}.
\newblock
\urldef\tempurl%
\url{https://doi.org/10.1145/564526.564544}
\showDOI{\tempurl}


\bibitem[Jänicke et~al\mbox{.}(2020)]%
        {visgap}
\bibfield{author}{\bibinfo{person}{Steffen Jänicke}, \bibinfo{person}{Pawandeep Kaur}, \bibinfo{person}{Pawel Kuzmicki}, {and} \bibinfo{person}{Johanna Schmidt}.} \bibinfo{year}{2020}\natexlab{}.
\newblock \showarticletitle{Participatory Visualization Design as an Approach to Minimize the Gap between Research and Application}.
\newblock
\urldef\tempurl%
\url{https://doi.org/10.2312/visgap.20201108}
\showDOI{\tempurl}


\bibitem[Kadijevich et~al\mbox{.}(2013)]%
        {kadijevich_angeli_schulte_2013}
\bibfield{author}{\bibinfo{person}{Djordje~M Kadijevich}, \bibinfo{person}{Charoula Angeli}, {and} \bibinfo{person}{Carsten Schulte}.} \bibinfo{year}{2013}\natexlab{}.
\newblock \bibinfo{booktitle}{\emph{Improving Computer Science Education}}.
\newblock \bibinfo{publisher}{Informa}.
\newblock
\urldef\tempurl%
\url{https://doi.org/10.4324/9780203078723}
\showDOI{\tempurl}


\bibitem[Kerzner et~al\mbox{.}(2019)]%
        {cvo}
\bibfield{author}{\bibinfo{person}{Ethan Kerzner}, \bibinfo{person}{Sarah Goodwin}, \bibinfo{person}{Jason Dykes}, \bibinfo{person}{Sara Jones}, {and} \bibinfo{person}{Miriah Meyer}.} \bibinfo{year}{2019}\natexlab{}.
\newblock \showarticletitle{A Framework for Creative Visualization-Opportunities Workshops}.
\newblock \bibinfo{journal}{\emph{IEEE Transactions on Visualization and Computer Graphics}} \bibinfo{volume}{25}, \bibinfo{number}{1} (\bibinfo{year}{2019}), \bibinfo{pages}{748--758}.
\newblock
\urldef\tempurl%
\url{https://doi.org/10.1109/TVCG.2018.2865241}
\showDOI{\tempurl}


\bibitem[Khot et~al\mbox{.}(2020)]%
        {khot2020shelfie}
\bibfield{author}{\bibinfo{person}{Rohit~Ashok Khot}, \bibinfo{person}{Larissa Hjorth}, {and} \bibinfo{person}{Florian Mueller}.} \bibinfo{year}{2020}\natexlab{}.
\newblock \showarticletitle{Shelfie: a framework for designing material representations of physical activity data}.
\newblock \bibinfo{journal}{\emph{ACM Transactions on Computer-Human Interaction (TOCHI)}} \bibinfo{volume}{27}, \bibinfo{number}{3} (\bibinfo{year}{2020}), \bibinfo{pages}{1--52}.
\newblock


\bibitem[Khot et~al\mbox{.}(2014)]%
        {Khot2014}
\bibfield{author}{\bibinfo{person}{Rohit~Ashok Khot}, \bibinfo{person}{Larissa Hjorth}, {and} \bibinfo{person}{Florian~“Floyd” Mueller}.} \bibinfo{year}{2014}\natexlab{}.
\newblock \showarticletitle{Understanding physical activity through 3D printed material artifacts}. In \bibinfo{booktitle}{\emph{Proceedings of the SIGCHI Conference on Human Factors in Computing Systems}} \emph{(\bibinfo{series}{CHI ’14})}. \bibinfo{publisher}{ACM}.
\newblock
\urldef\tempurl%
\url{https://doi.org/10.1145/2556288.2557144}
\showDOI{\tempurl}


\bibitem[Khowaja et~al\mbox{.}(2022)]%
        {activis}
\bibfield{author}{\bibinfo{person}{Kamran Khowaja}, \bibinfo{person}{Wafa~Waheeda Syed}, \bibinfo{person}{Meghna Singh}, \bibinfo{person}{Shahrad Taheri}, \bibinfo{person}{Odette Chagoury}, \bibinfo{person}{Dena Al-Thani}, {and} \bibinfo{person}{Micha{\"e}l Aupetit}.} \bibinfo{year}{2022}\natexlab{}.
\newblock \showarticletitle{A Participatory Design Approach to Develop Visualization of Wearable Actigraphy Data for Health Care Professionals: Case Study in Qatar}.
\newblock \bibinfo{journal}{\emph{JMIR Hum Factors}} \bibinfo{volume}{9}, \bibinfo{number}{2} (\bibinfo{date}{8 Apr} \bibinfo{year}{2022}), \bibinfo{pages}{e25880}.
\newblock
\showISSN{2292-9495}
\urldef\tempurl%
\url{https://doi.org/10.2196/25880}
\showDOI{\tempurl}


\bibitem[Kim et~al\mbox{.}(2024)]%
        {kim2024opportunities}
\bibfield{author}{\bibinfo{person}{Taewook Kim}, \bibinfo{person}{Hyeok Kim}, \bibinfo{person}{Angela Roberts}, \bibinfo{person}{Maia Jacobs}, {and} \bibinfo{person}{Matthew Kay}.} \bibinfo{year}{2024}\natexlab{}.
\newblock \bibinfo{title}{Opportunities in Mental Health Support for Informal Dementia Caregivers Suffering from Verbal Agitation}.
\newblock
\newblock


\bibitem[Kosara and Mackinlay(2013)]%
        {Kosara2013}
\bibfield{author}{\bibinfo{person}{Robert Kosara} {and} \bibinfo{person}{Jock Mackinlay}.} \bibinfo{year}{2013}\natexlab{}.
\newblock \showarticletitle{Storytelling: The Next Step for Visualization}.
\newblock \bibinfo{journal}{\emph{Computer}} \bibinfo{volume}{46}, \bibinfo{number}{5} (\bibinfo{date}{May} \bibinfo{year}{2013}), \bibinfo{pages}{44–50}.
\newblock
\showISSN{0018-9162}
\urldef\tempurl%
\url{https://doi.org/10.1109/mc.2013.36}
\showDOI{\tempurl}


\bibitem[Koushik et~al\mbox{.}(2019)]%
        {Koushik2019}
\bibfield{author}{\bibinfo{person}{Varsha Koushik}, \bibinfo{person}{Darren Guinness}, {and} \bibinfo{person}{Shaun~K. Kane}.} \bibinfo{year}{2019}\natexlab{}.
\newblock \showarticletitle{StoryBlocks: A Tangible Programming Game To Create Accessible Audio Stories}. In \bibinfo{booktitle}{\emph{Proceedings of the 2019 CHI Conference on Human Factors in Computing Systems}} \emph{(\bibinfo{series}{CHI ’19})}. \bibinfo{publisher}{ACM}.
\newblock
\urldef\tempurl%
\url{https://doi.org/10.1145/3290605.3300722}
\showDOI{\tempurl}


\bibitem[Koushik and Kane(2022)]%
        {koushik2022towards}
\bibfield{author}{\bibinfo{person}{Varsha Koushik} {and} \bibinfo{person}{Shaun~K Kane}.} \bibinfo{year}{2022}\natexlab{}.
\newblock \showarticletitle{Towards augmented reality coaching for daily routines: Participatory design with individuals with cognitive disabilities and their caregivers}.
\newblock \bibinfo{journal}{\emph{International Journal of Human-Computer Studies}}  \bibinfo{volume}{165} (\bibinfo{year}{2022}), \bibinfo{pages}{102862}.
\newblock


\bibitem[Lan et~al\mbox{.}(2024)]%
        {Lan2024}
\bibfield{author}{\bibinfo{person}{Xingyu Lan}, \bibinfo{person}{Yanqiu Wu}, {and} \bibinfo{person}{Nan Cao}.} \bibinfo{year}{2024}\natexlab{}.
\newblock \showarticletitle{Affective Visualization Design: Leveraging the Emotional Impact of Data}.
\newblock \bibinfo{journal}{\emph{IEEE Transactions on Visualization and Computer Graphics}} \bibinfo{volume}{30}, \bibinfo{number}{1} (\bibinfo{date}{Jan.} \bibinfo{year}{2024}), \bibinfo{pages}{1–11}.
\newblock
\showISSN{2160-9306}
\urldef\tempurl%
\url{https://doi.org/10.1109/tvcg.2023.3327385}
\showDOI{\tempurl}


\bibitem[Lee et~al\mbox{.}(2020)]%
        {lee2020reaching}
\bibfield{author}{\bibinfo{person}{Bongshin Lee}, \bibinfo{person}{Eun~Kyoung Choe}, \bibinfo{person}{Petra Isenberg}, \bibinfo{person}{Kim Marriott}, {and} \bibinfo{person}{John Stasko}.} \bibinfo{year}{2020}\natexlab{}.
\newblock \showarticletitle{Reaching broader audiences with data visualization}.
\newblock \bibinfo{journal}{\emph{IEEE Computer Graphics and Applications}} \bibinfo{volume}{40}, \bibinfo{number}{2} (\bibinfo{year}{2020}), \bibinfo{pages}{82--90}.
\newblock


\bibitem[Lee et~al\mbox{.}(2013)]%
        {lee2013sketchstory}
\bibfield{author}{\bibinfo{person}{Bongshin Lee}, \bibinfo{person}{Rubaiat~Habib Kazi}, {and} \bibinfo{person}{Greg Smith}.} \bibinfo{year}{2013}\natexlab{}.
\newblock \showarticletitle{SketchStory: Telling more engaging stories with data through freeform sketching}.
\newblock \bibinfo{journal}{\emph{IEEE transactions on visualization and computer graphics}} \bibinfo{volume}{19}, \bibinfo{number}{12} (\bibinfo{year}{2013}), \bibinfo{pages}{2416--2425}.
\newblock


\bibitem[Lee et~al\mbox{.}(2024)]%
        {lee2024inclusive}
\bibfield{author}{\bibinfo{person}{Bongshin Lee}, \bibinfo{person}{Kim Marriott}, \bibinfo{person}{Danielle Szafir}, {and} \bibinfo{person}{Gerhard Weber}.} \bibinfo{year}{2024}\natexlab{}.
\newblock \showarticletitle{Inclusive Data Visualization (Dagstuhl Seminar 23252)}.
\newblock  (\bibinfo{year}{2024}).
\newblock


\bibitem[Lee et~al\mbox{.}(2018)]%
        {Lee2018}
\bibfield{author}{\bibinfo{person}{Jung~Min Lee}, \bibinfo{person}{Jongsoo Baek}, {and} \bibinfo{person}{Da~Young Ju}.} \bibinfo{year}{2018}\natexlab{}.
\newblock \showarticletitle{Anthropomorphic Design: Emotional Perception for Deformable Object}.
\newblock \bibinfo{journal}{\emph{Frontiers in Psychology}}  \bibinfo{volume}{9} (\bibinfo{date}{Oct.} \bibinfo{year}{2018}).
\newblock
\showISSN{1664-1078}
\urldef\tempurl%
\url{https://doi.org/10.3389/fpsyg.2018.01829}
\showDOI{\tempurl}


\bibitem[Lee and Hong(2017)]%
        {Lee2017}
\bibfield{author}{\bibinfo{person}{Kwangyoung Lee} {and} \bibinfo{person}{Hwajung Hong}.} \bibinfo{year}{2017}\natexlab{}.
\newblock \showarticletitle{Designing for Self-Tracking of Emotion and Experience with Tangible Modality}. In \bibinfo{booktitle}{\emph{Proceedings of the 2017 Conference on Designing Interactive Systems}} \emph{(\bibinfo{series}{DIS ’17})}. \bibinfo{publisher}{ACM}.
\newblock
\urldef\tempurl%
\url{https://doi.org/10.1145/3064663.3064697}
\showDOI{\tempurl}


\bibitem[Lee et~al\mbox{.}(2017)]%
        {lee_vlat_2017}
\bibfield{author}{\bibinfo{person}{Sukwon Lee}, \bibinfo{person}{Sung-Hee Kim}, {and} \bibinfo{person}{Bum~Chul Kwon}.} \bibinfo{year}{2017}\natexlab{}.
\newblock \showarticletitle{{VLAT}: {Development} of a {Visualization} {Literacy} {Assessment} {Test}}.
\newblock \bibinfo{journal}{\emph{IEEE Transactions on Visualization and Computer Graphics}} \bibinfo{volume}{23}, \bibinfo{number}{1} (\bibinfo{year}{2017}), \bibinfo{pages}{551--560}.
\newblock


\bibitem[Lee-Robbins and Adar(2022)]%
        {LeeRobbins2022}
\bibfield{author}{\bibinfo{person}{Elsie Lee-Robbins} {and} \bibinfo{person}{Eytan Adar}.} \bibinfo{year}{2022}\natexlab{}.
\newblock \showarticletitle{Affective Learning Objectives for Communicative Visualizations}.
\newblock \bibinfo{journal}{\emph{IEEE Transactions on Visualization and Computer Graphics}} (\bibinfo{year}{2022}), \bibinfo{pages}{1–11}.
\newblock
\showISSN{2160-9306}
\urldef\tempurl%
\url{https://doi.org/10.1109/tvcg.2022.3209500}
\showDOI{\tempurl}


\bibitem[Legaki and Hamari(2020)]%
        {gamedata}
\bibfield{author}{\bibinfo{person}{Zampeta Legaki} {and} \bibinfo{person}{Juho Hamari}.} \bibinfo{year}{2020}\natexlab{}.
\newblock \showarticletitle{Gamification in statistics education: A literature review}. In \bibinfo{booktitle}{\emph{GamiFIN Conference 2020}} \emph{(\bibinfo{series}{CEUR workshop proceedings})}, \bibfield{editor}{\bibinfo{person}{Jonna Koivisto}, \bibinfo{person}{Mila Buji{\'c}}, {and} \bibinfo{person}{Juho Hamari}} (Eds.). \bibinfo{publisher}{CEUR-WS}, \bibinfo{pages}{41--51}.
\newblock
\newblock
\shownote{JUFOID=53269; International GamiFIN Conference ; Conference date: 01-01-1900}.


\bibitem[Liem et~al\mbox{.}(2020)]%
        {attitude}
\bibfield{author}{\bibinfo{person}{Johannes Liem}, \bibinfo{person}{Charles Perin}, {and} \bibinfo{person}{Jo Wood}.} \bibinfo{year}{2020}\natexlab{}.
\newblock \showarticletitle{Structure and Empathy in Visual Data Storytelling: Evaluating their Influence on Attitude}.
\newblock \bibinfo{journal}{\emph{Computer Graphics Forum}}  \bibinfo{volume}{39} (\bibinfo{date}{06} \bibinfo{year}{2020}).
\newblock
\urldef\tempurl%
\url{https://doi.org/10.1111/cgf.13980}
\showDOI{\tempurl}


\bibitem[Lin et~al\mbox{.}(2023)]%
        {datahunches}
\bibfield{author}{\bibinfo{person}{Haihan Lin}, \bibinfo{person}{Derya Akbaba}, \bibinfo{person}{Miriah Meyer}, {and} \bibinfo{person}{Alexander Lex}.} \bibinfo{year}{2023}\natexlab{}.
\newblock \showarticletitle{Data Hunches: Incorporating Personal Knowledge into Visualizations}.
\newblock \bibinfo{journal}{\emph{IEEE Transactions on Visualization and Computer Graphics}} \bibinfo{volume}{29}, \bibinfo{number}{1} (\bibinfo{year}{2023}), \bibinfo{pages}{504--514}.
\newblock
\urldef\tempurl%
\url{https://doi.org/10.1109/TVCG.2022.3209451}
\showDOI{\tempurl}


\bibitem[Lindstr\"{o}m et~al\mbox{.}(2023)]%
        {Lindstrm2023}
\bibfield{author}{\bibinfo{person}{Esther~R. Lindstr\"{o}m}, \bibinfo{person}{Emma Fisher}, \bibinfo{person}{Megan Cook}, \bibinfo{person}{Mariangela Perrella}, \bibinfo{person}{Kimberly~A. McFadden}, \bibinfo{person}{Rui Chen}, {and} \bibinfo{person}{Mohammad~Bahadori Fallah}.} \bibinfo{year}{2023}\natexlab{}.
\newblock \showarticletitle{An observation study of mathematics instruction for students with IDD in grades K-2}.
\newblock \bibinfo{journal}{\emph{Research in Developmental Disabilities}}  \bibinfo{volume}{141} (\bibinfo{date}{Oct.} \bibinfo{year}{2023}), \bibinfo{pages}{104591}.
\newblock
\showISSN{0891-4222}
\urldef\tempurl%
\url{https://doi.org/10.1016/j.ridd.2023.104591}
\showDOI{\tempurl}


\bibitem[Lloyd-Esenkaya et~al\mbox{.}(2020)]%
        {lloyd2020multisensory}
\bibfield{author}{\bibinfo{person}{Tayfun Lloyd-Esenkaya}, \bibinfo{person}{Vanessa Lloyd-Esenkaya}, \bibinfo{person}{Eamonn O’Neill}, {and} \bibinfo{person}{Michael~J Proulx}.} \bibinfo{year}{2020}\natexlab{}.
\newblock \showarticletitle{Multisensory inclusive design with sensory substitution}.
\newblock \bibinfo{journal}{\emph{Cognitive Research: Principles and Implications}} \bibinfo{volume}{5}, \bibinfo{number}{1} (\bibinfo{year}{2020}), \bibinfo{pages}{37}.
\newblock


\bibitem[Lund(2022)]%
        {hiphop}
\bibfield{author}{\bibinfo{person}{Brady~D. Lund}.} \bibinfo{year}{2022}\natexlab{}.
\newblock \showarticletitle{The Art of (Data) Storytelling: Hip Hop Innovation and Bringing a Social Justice Mindset to Data Science and Visualization}.
\newblock \bibinfo{journal}{\emph{The International Journal of Information, Diversity, \& Inclusion}} \bibinfo{volume}{6}, \bibinfo{number}{1/2} (\bibinfo{year}{2022}), \bibinfo{pages}{31--41}.
\newblock
\showISSN{25743430}
\urldef\tempurl%
\url{https://www.jstor.org/stable/48665362}
\showURL{%
\tempurl}


\bibitem[Lundgard et~al\mbox{.}(2019)]%
        {lundgard2019sociotechnical}
\bibfield{author}{\bibinfo{person}{Alan Lundgard}, \bibinfo{person}{Crystal Lee}, {and} \bibinfo{person}{Arvind Satyanarayan}.} \bibinfo{year}{2019}\natexlab{}.
\newblock \showarticletitle{Sociotechnical considerations for accessible visualization design}. In \bibinfo{booktitle}{\emph{2019 IEEE Visualization Conference (VIS)}}. IEEE, \bibinfo{pages}{16--20}.
\newblock


\bibitem[Marriott et~al\mbox{.}(2021)]%
        {marriott2021inclusive}
\bibfield{author}{\bibinfo{person}{Kim Marriott}, \bibinfo{person}{Bongshin Lee}, \bibinfo{person}{Matthew Butler}, \bibinfo{person}{Ed Cutrell}, \bibinfo{person}{Kirsten Ellis}, \bibinfo{person}{Cagatay Goncu}, \bibinfo{person}{Marti Hearst}, \bibinfo{person}{Kathleen McCoy}, {and} \bibinfo{person}{Danielle~Albers Szafir}.} \bibinfo{year}{2021}\natexlab{}.
\newblock \showarticletitle{Inclusive data visualization for people with disabilities: a call to action}.
\newblock \bibinfo{journal}{\emph{Interactions}} \bibinfo{volume}{28}, \bibinfo{number}{3} (\bibinfo{year}{2021}), \bibinfo{pages}{47--51}.
\newblock


\bibitem[McCambridge et~al\mbox{.}(2014)]%
        {McCambridge2014}
\bibfield{author}{\bibinfo{person}{Jim McCambridge}, \bibinfo{person}{John Witton}, {and} \bibinfo{person}{Diana~R. Elbourne}.} \bibinfo{year}{2014}\natexlab{}.
\newblock \showarticletitle{Systematic review of the Hawthorne effect: New concepts are needed to study research participation effects}.
\newblock \bibinfo{journal}{\emph{Journal of Clinical Epidemiology}} \bibinfo{volume}{67}, \bibinfo{number}{3} (\bibinfo{date}{March} \bibinfo{year}{2014}), \bibinfo{pages}{267–277}.
\newblock
\showISSN{0895-4356}
\urldef\tempurl%
\url{https://doi.org/10.1016/j.jclinepi.2013.08.015}
\showDOI{\tempurl}


\bibitem[Mcclure et~al\mbox{.}(2009)]%
        {emotionaldysregulation}
\bibfield{author}{\bibinfo{person}{Kelly Mcclure}, \bibinfo{person}{Jacqueline Halpern}, \bibinfo{person}{Pamela Wolper}, {and} \bibinfo{person}{John Donahue}.} \bibinfo{year}{2009}\natexlab{}.
\newblock \showarticletitle{Emotion Regulation and Intellectual Disability}.
\newblock \bibinfo{journal}{\emph{Journal of Developmental Disabilities}}  \bibinfo{volume}{15} (\bibinfo{date}{01} \bibinfo{year}{2009}).
\newblock


\bibitem[mohammadyari et~al\mbox{.}(2021)]%
        {Rahgoi}
\bibfield{author}{\bibinfo{person}{pari mohammadyari}, \bibinfo{person}{abolfazl Rahgoi}, \bibinfo{person}{Masoud Fallahi-Khoshknab}, {and} \bibinfo{person}{Mohsen~and vahedi}.} \bibinfo{year}{2021}\natexlab{}.
\newblock \showarticletitle{The Effect of Storytelling on Visual and Auditory Attention andConcentration in Children with Autism Spectrum Disorders}.
\newblock \bibinfo{journal}{\emph{Iranian Journal of Rehabilitation Research in Nursing}} \bibinfo{volume}{7}, \bibinfo{number}{4} (\bibinfo{year}{2021}).
\newblock
\urldef\tempurl%
\url{https://doi.org/10.22034/IJRN.7.4.1}
\showDOI{\tempurl}
\showeprint{http://ijrn.ir/article-1-642-en.pdf}


\bibitem[Molina~León and Breiter(2020)]%
        {co-creation}
\bibfield{author}{\bibinfo{person}{Gabriela Molina~León} {and} \bibinfo{person}{Andreas Breiter}.} \bibinfo{year}{2020}\natexlab{}.
\newblock \showarticletitle{{Co-creating Visualizations: A First Evaluation with Social Science Researchers}}.
\newblock \bibinfo{journal}{\emph{Computer Graphics Forum}} (\bibinfo{year}{2020}).
\newblock
\showISSN{1467-8659}
\urldef\tempurl%
\url{https://doi.org/10.1111/cgf.13981}
\showDOI{\tempurl}


\bibitem[Neate et~al\mbox{.}(2019)]%
        {Neate2019}
\bibfield{author}{\bibinfo{person}{Timothy Neate}, \bibinfo{person}{Aikaterini Bourazeri}, \bibinfo{person}{Abi Roper}, \bibinfo{person}{Simone Stumpf}, {and} \bibinfo{person}{Stephanie Wilson}.} \bibinfo{year}{2019}\natexlab{}.
\newblock \showarticletitle{Co-Created Personas: Engaging and Empowering Users with Diverse Needs Within the Design Process}. In \bibinfo{booktitle}{\emph{Proceedings of the 2019 CHI Conference on Human Factors in Computing Systems}} \emph{(\bibinfo{series}{CHI ’19})}. \bibinfo{publisher}{ACM}.
\newblock
\urldef\tempurl%
\url{https://doi.org/10.1145/3290605.3300880}
\showDOI{\tempurl}


\bibitem[Nurain and Chung(2023)]%
        {Nurain2023}
\bibfield{author}{\bibinfo{person}{Novia Nurain} {and} \bibinfo{person}{Chia-Fang Chung}.} \bibinfo{year}{2023}\natexlab{}.
\newblock \showarticletitle{“I left my legacy, told my story”: Understanding Older Adults’ Tracking Practices to Promote Active Aging}. In \bibinfo{booktitle}{\emph{Proceedings of the 2023 ACM Designing Interactive Systems Conference}} \emph{(\bibinfo{series}{DIS ’23})}. \bibinfo{publisher}{ACM}.
\newblock
\urldef\tempurl%
\url{https://doi.org/10.1145/3563657.3596083}
\showDOI{\tempurl}


\bibitem[Payne et~al\mbox{.}(2021)]%
        {Payne2021}
\bibfield{author}{\bibinfo{person}{William~Christopher Payne}, \bibinfo{person}{Yoav Bergner}, \bibinfo{person}{Mary~Etta West}, \bibinfo{person}{Carlie Charp}, \bibinfo{person}{R.~Benjamin Shapiro}, \bibinfo{person}{Danielle~Albers Szafir}, \bibinfo{person}{Edd~V. Taylor}, {and} \bibinfo{person}{Kayla DesPortes}.} \bibinfo{year}{2021}\natexlab{}.
\newblock \showarticletitle{danceON: Culturally Responsive Creative Computing}. In \bibinfo{booktitle}{\emph{Proceedings of the 2021 CHI Conference on Human Factors in Computing Systems}} \emph{(\bibinfo{series}{CHI ’21})}. \bibinfo{publisher}{ACM}.
\newblock
\urldef\tempurl%
\url{https://doi.org/10.1145/3411764.3445149}
\showDOI{\tempurl}


\bibitem[Peck et~al\mbox{.}(2019)]%
        {dataispersonal}
\bibfield{author}{\bibinfo{person}{Evan~M. Peck}, \bibinfo{person}{Sofia~E. Ayuso}, {and} \bibinfo{person}{Omar El-Etr}.} \bibinfo{year}{2019}\natexlab{}.
\newblock \showarticletitle{Data is Personal: Attitudes and Perceptions of Data Visualization in Rural Pennsylvania}. In \bibinfo{booktitle}{\emph{Proceedings of the 2019 CHI Conference on Human Factors in Computing Systems}} (Glasgow, Scotland Uk) \emph{(\bibinfo{series}{CHI '19})}. \bibinfo{publisher}{Association for Computing Machinery}, \bibinfo{address}{New York, NY, USA}, \bibinfo{pages}{1–12}.
\newblock
\showISBNx{9781450359702}
\urldef\tempurl%
\url{https://doi.org/10.1145/3290605.3300474}
\showDOI{\tempurl}


\bibitem[Polat et~al\mbox{.}(2019)]%
        {Polat2019}
\bibfield{author}{\bibinfo{person}{Elif Polat}, \bibinfo{person}{Kursat Cagiltay}, {and} \bibinfo{person}{Necdet Karasu}.} \bibinfo{year}{2019}\natexlab{}.
\newblock \bibinfo{booktitle}{\emph{Tangible Objects and Mobile Technology: Interactive Learning Environments for Students with Learning Disabilities}}.
\newblock \bibinfo{publisher}{Springer Singapore}, \bibinfo{pages}{635–653}.
\newblock
\showISBNx{9789811327667}
\urldef\tempurl%
\url{https://doi.org/10.1007/978-981-13-2766-7_119}
\showDOI{\tempurl}


\bibitem[Rathnayake et~al\mbox{.}(2021)]%
        {rathnayake2021co}
\bibfield{author}{\bibinfo{person}{Sarath Rathnayake}, \bibinfo{person}{Wendy Moyle}, \bibinfo{person}{Cindy Jones}, {and} \bibinfo{person}{Pauline Calleja}.} \bibinfo{year}{2021}\natexlab{}.
\newblock \showarticletitle{Co-design of an mHealth application for family caregivers of people with dementia to address functional disability care needs}.
\newblock \bibinfo{journal}{\emph{Informatics for Health and Social Care}} \bibinfo{volume}{46}, \bibinfo{number}{1} (\bibinfo{year}{2021}), \bibinfo{pages}{1--17}.
\newblock


\bibitem[Ravihansa~Rajapakse and Sitbon(2021)]%
        {respectful}
\bibfield{author}{\bibinfo{person}{Margot~Brereton Ravihansa~Rajapakse} {and} \bibinfo{person}{Laurianne Sitbon}.} \bibinfo{year}{2021}\natexlab{}.
\newblock \showarticletitle{A respectful design approach to facilitate codesign with people with cognitive or sensory impairments and makers}.
\newblock \bibinfo{journal}{\emph{CoDesign}} \bibinfo{volume}{17}, \bibinfo{number}{2} (\bibinfo{year}{2021}), \bibinfo{pages}{159--187}.
\newblock
\urldef\tempurl%
\url{https://doi.org/10.1080/15710882.2019.1612442}
\showDOI{\tempurl}
\showeprint{https://doi.org/10.1080/15710882.2019.1612442}


\bibitem[Ren et~al\mbox{.}(2017)]%
        {chartaccent}
\bibfield{author}{\bibinfo{person}{Donghao Ren}, \bibinfo{person}{Matthew Brehmer}, \bibinfo{person}{Bongshin Lee}, \bibinfo{person}{Tobias Höllerer}, {and} \bibinfo{person}{Eun~Kyoung Choe}.} \bibinfo{year}{2017}\natexlab{}.
\newblock \showarticletitle{ChartAccent: Annotation for data-driven storytelling}. In \bibinfo{booktitle}{\emph{2017 IEEE Pacific Visualization Symposium (PacificVis)}}. \bibinfo{pages}{230--239}.
\newblock
\showISSN{2165-8773}
\urldef\tempurl%
\url{https://doi.org/10.1109/PACIFICVIS.2017.8031599}
\showDOI{\tempurl}


\bibitem[Roberg\'{e} and Carlson(1997)]%
        {broadening}
\bibfield{author}{\bibinfo{person}{James Roberg\'{e}} {and} \bibinfo{person}{C.~R. Carlson}.} \bibinfo{year}{1997}\natexlab{}.
\newblock \showarticletitle{Broadening the computer science curriculum}.
\newblock \bibinfo{journal}{\emph{SIGCSE Bull.}} \bibinfo{volume}{29}, \bibinfo{number}{1} (\bibinfo{date}{mar} \bibinfo{year}{1997}), \bibinfo{pages}{320–324}.
\newblock
\showISSN{0097-8418}
\urldef\tempurl%
\url{https://doi.org/10.1145/268085.268206}
\showDOI{\tempurl}


\bibitem[Robinson et~al\mbox{.}(2014)]%
        {storytellingtv}
\bibfield{author}{\bibinfo{person}{Susan~J. Robinson}, \bibinfo{person}{Graceline Williams}, \bibinfo{person}{Aman Parnami}, \bibinfo{person}{Jinhyun Kim}, \bibinfo{person}{Emmett McGregor}, \bibinfo{person}{Dana Chandler}, {and} \bibinfo{person}{Ali Mazalek}.} \bibinfo{year}{2014}\natexlab{}.
\newblock \showarticletitle{Storied numbers: supporting media-rich data storytelling for television}. In \bibinfo{booktitle}{\emph{Proceedings of the ACM International Conference on Interactive Experiences for TV and Online Video}} (Newcastle Upon Tyne, United Kingdom) \emph{(\bibinfo{series}{TVX '14})}. \bibinfo{publisher}{Association for Computing Machinery}, \bibinfo{address}{New York, NY, USA}, \bibinfo{pages}{123–130}.
\newblock
\showISBNx{9781450328388}
\urldef\tempurl%
\url{https://doi.org/10.1145/2602299.2602308}
\showDOI{\tempurl}


\bibitem[Sakamoto et~al\mbox{.}(2022)]%
        {affectstorytelling}
\bibfield{author}{\bibinfo{person}{Yumiko Sakamoto}, \bibinfo{person}{Samar Sallam}, \bibinfo{person}{Aaron Salo}, \bibinfo{person}{Jason Leboe-McGowan}, {and} \bibinfo{person}{Pourang Irani}.} \bibinfo{year}{2022}\natexlab{}.
\newblock \showarticletitle{Persuasive Data Storytelling with a Data Video during Covid-19 Infodemic: Affective Pathway to Influence the Users' Perception about Contact Tracing Apps in less than 6 Minutes}. In \bibinfo{booktitle}{\emph{2022 IEEE 15th Pacific Visualization Symposium (PacificVis)}}. \bibinfo{pages}{176--180}.
\newblock
\urldef\tempurl%
\url{https://doi.org/10.1109/PacificVis53943.2022.00028}
\showDOI{\tempurl}


\bibitem[Samson et~al\mbox{.}(2014)]%
        {Samson2014}
\bibfield{author}{\bibinfo{person}{Andrea~C. Samson}, \bibinfo{person}{Antonio~Y. Hardan}, \bibinfo{person}{Rebecca~W. Podell}, \bibinfo{person}{Jennifer~M. Phillips}, {and} \bibinfo{person}{James~J. Gross}.} \bibinfo{year}{2014}\natexlab{}.
\newblock \showarticletitle{Emotion Regulation in Children and Adolescents With Autism Spectrum Disorder}.
\newblock \bibinfo{journal}{\emph{Autism Research}} \bibinfo{volume}{8}, \bibinfo{number}{1} (\bibinfo{date}{May} \bibinfo{year}{2014}), \bibinfo{pages}{9–18}.
\newblock
\showISSN{1939-3806}
\urldef\tempurl%
\url{https://doi.org/10.1002/aur.1387}
\showDOI{\tempurl}


\bibitem[Sanders and Stappers(2008)]%
        {codesign}
\bibfield{author}{\bibinfo{person}{Elizabeth Sanders} {and} \bibinfo{person}{Pieter~Jan Stappers}.} \bibinfo{year}{2008}\natexlab{}.
\newblock \showarticletitle{Co-creation and the New Landscapes of Design}.
\newblock \bibinfo{journal}{\emph{CoDesign}}  \bibinfo{volume}{4} (\bibinfo{date}{03} \bibinfo{year}{2008}), \bibinfo{pages}{5--18}.
\newblock
\urldef\tempurl%
\url{https://doi.org/10.1080/15710880701875068}
\showDOI{\tempurl}


\bibitem[Saridaki and Meimaris(2018)]%
        {saridaki2018digital}
\bibfield{author}{\bibinfo{person}{Maria Saridaki} {and} \bibinfo{person}{Michalis Meimaris}.} \bibinfo{year}{2018}\natexlab{}.
\newblock \showarticletitle{Digital Storytelling for the empowerment of people with intellectual disabilities}. In \bibinfo{booktitle}{\emph{Proceedings of the 8th international conference on software development and technologies for enhancing accessibility and fighting info-exclusion}}. \bibinfo{pages}{161--164}.
\newblock


\bibitem[Sarmiento-Pelayo(2015)]%
        {sarmiento2015co}
\bibfield{author}{\bibinfo{person}{Martha~Patricia Sarmiento-Pelayo}.} \bibinfo{year}{2015}\natexlab{}.
\newblock \showarticletitle{Co-design: A central approach to the inclusion of people with disabilities}.
\newblock \bibinfo{journal}{\emph{Revista de la Facultad de Medicina}}  \bibinfo{volume}{63} (\bibinfo{year}{2015}), \bibinfo{pages}{149--154}.
\newblock


\bibitem[Segel and Heer(2010)]%
        {segel2010narrative}
\bibfield{author}{\bibinfo{person}{Edward Segel} {and} \bibinfo{person}{Jeffrey Heer}.} \bibinfo{year}{2010}\natexlab{}.
\newblock \showarticletitle{Narrative visualization: Telling stories with data}.
\newblock \bibinfo{journal}{\emph{IEEE transactions on visualization and computer graphics}} \bibinfo{volume}{16}, \bibinfo{number}{6} (\bibinfo{year}{2010}), \bibinfo{pages}{1139--1148}.
\newblock


\bibitem[Sharif et~al\mbox{.}(2022)]%
        {voxlens}
\bibfield{author}{\bibinfo{person}{Ather Sharif}, \bibinfo{person}{Olivia~H. Wang}, \bibinfo{person}{Alida~T. Muongchan}, \bibinfo{person}{Katharina Reinecke}, {and} \bibinfo{person}{Jacob~O. Wobbrock}.} \bibinfo{year}{2022}\natexlab{}.
\newblock \showarticletitle{VoxLens: Making Online Data Visualizations Accessible with an Interactive JavaScript Plug-In}. In \bibinfo{booktitle}{\emph{Proceedings of the 2022 CHI Conference on Human Factors in Computing Systems}} (New Orleans, LA, USA) \emph{(\bibinfo{series}{CHI '22})}. \bibinfo{publisher}{Association for Computing Machinery}, \bibinfo{address}{New York, NY, USA}, Article \bibinfo{articleno}{478}, \bibinfo{numpages}{19}~pages.
\newblock
\showISBNx{9781450391573}
\urldef\tempurl%
\url{https://doi.org/10.1145/3491102.3517431}
\showDOI{\tempurl}


\bibitem[Snyder et~al\mbox{.}(2020)]%
        {cancer}
\bibfield{author}{\bibinfo{person}{Lauren Snyder}, \bibinfo{person}{Ayan~Anandkumar Saraf}, \bibinfo{person}{Reggie Casanova-Perez}, \bibinfo{person}{Sarah~E. Connor}, \bibinfo{person}{Sheba George}, \bibinfo{person}{Amelia Wang}, \bibinfo{person}{Darwin Jones}, \bibinfo{person}{Georgina Mendoza}, \bibinfo{person}{John~L. Gore}, \bibinfo{person}{Mark~S. Litwin}, {and} \bibinfo{person}{Andrea Hartzler}.} \bibinfo{year}{2020}\natexlab{}.
\newblock \showarticletitle{Visualization Co-Design with Prostate Cancer Survivors who have Limited Graph Literacy}. In \bibinfo{booktitle}{\emph{2020 Workshop on Visual Analytics in Healthcare (VAHC)}}. \bibinfo{pages}{17--23}.
\newblock
\urldef\tempurl%
\url{https://doi.org/10.1109/VAHC53729.2020.00009}
\showDOI{\tempurl}


\bibitem[Soares~Guedes et~al\mbox{.}(2022)]%
        {designwithid}
\bibfield{author}{\bibinfo{person}{Leandro Soares~Guedes}, \bibinfo{person}{Ryan~Colin Gibson}, \bibinfo{person}{Kirsten Ellis}, \bibinfo{person}{Laurianne Sitbon}, {and} \bibinfo{person}{Monica Landoni}.} \bibinfo{year}{2022}\natexlab{}.
\newblock \showarticletitle{Designing with and for People with Intellectual Disabilities}. In \bibinfo{booktitle}{\emph{Proceedings of the 24th International ACM SIGACCESS Conference on Computers and Accessibility}} \emph{(\bibinfo{series}{ASSETS '22})}. \bibinfo{publisher}{Association for Computing Machinery}, \bibinfo{address}{New York, NY, USA}, Article \bibinfo{articleno}{106}, \bibinfo{numpages}{6}~pages.
\newblock
\showISBNx{9781450392587}
\urldef\tempurl%
\url{https://doi.org/10.1145/3517428.3550406}
\showDOI{\tempurl}


\bibitem[Spinuzzi(2005)]%
        {spinuzzi2005methodology}
\bibfield{author}{\bibinfo{person}{Clay Spinuzzi}.} \bibinfo{year}{2005}\natexlab{}.
\newblock \showarticletitle{The methodology of participatory design}.
\newblock \bibinfo{journal}{\emph{Technical communication}} \bibinfo{volume}{52}, \bibinfo{number}{2} (\bibinfo{year}{2005}), \bibinfo{pages}{163--174}.
\newblock


\bibitem[Stieff et~al\mbox{.}(2020)]%
        {Stieff2020}
\bibfield{author}{\bibinfo{person}{Mike Stieff}, \bibinfo{person}{Stephanie Werner}, \bibinfo{person}{Dane DeSutter}, \bibinfo{person}{Steve Franconeri}, {and} \bibinfo{person}{Mary Hegarty}.} \bibinfo{year}{2020}\natexlab{}.
\newblock \showarticletitle{Visual chunking as a strategy for spatial thinking in STEM}.
\newblock \bibinfo{journal}{\emph{Cognitive Research: Principles and Implications}} \bibinfo{volume}{5}, \bibinfo{number}{1} (\bibinfo{date}{April} \bibinfo{year}{2020}).
\newblock
\showISSN{2365-7464}
\urldef\tempurl%
\url{https://doi.org/10.1186/s41235-020-00217-6}
\showDOI{\tempurl}


\bibitem[Thoft et~al\mbox{.}(2020)]%
        {voice}
\bibfield{author}{\bibinfo{person}{Diana~Schack Thoft}, \bibinfo{person}{Michelle Pyer}, \bibinfo{person}{Anders Horsbøl}, {and} \bibinfo{person}{Jacqueline Parkes}.} \bibinfo{year}{2020}\natexlab{}.
\newblock \showarticletitle{The Balanced Participation Model: Sharing opportunities for giving people with early-stage dementia a voice in research}.
\newblock \bibinfo{journal}{\emph{Dementia}} \bibinfo{volume}{19}, \bibinfo{number}{7} (\bibinfo{year}{2020}), \bibinfo{pages}{2294--2313}.
\newblock
\urldef\tempurl%
\url{https://doi.org/10.1177/1471301218820208}
\showDOI{\tempurl}
\newblock
\shownote{PMID: 30587030}.


\bibitem[Thompson et~al\mbox{.}(2023)]%
        {chartreader}
\bibfield{author}{\bibinfo{person}{John~R Thompson}, \bibinfo{person}{Jesse~J Martinez}, \bibinfo{person}{Alper Sarikaya}, \bibinfo{person}{Edward Cutrell}, {and} \bibinfo{person}{Bongshin Lee}.} \bibinfo{year}{2023}\natexlab{}.
\newblock \showarticletitle{Chart Reader: Accessible Visualization Experiences Designed with Screen Reader Users}. In \bibinfo{booktitle}{\emph{Proceedings of the 2023 CHI Conference on Human Factors in Computing Systems}} (Hamburg, Germany) \emph{(\bibinfo{series}{CHI '23})}. \bibinfo{publisher}{Association for Computing Machinery}, \bibinfo{address}{New York, NY, USA}, Article \bibinfo{articleno}{802}, \bibinfo{numpages}{18}~pages.
\newblock
\showISBNx{9781450394215}
\urldef\tempurl%
\url{https://doi.org/10.1145/3544548.3581186}
\showDOI{\tempurl}


\bibitem[Thudt et~al\mbox{.}(2016)]%
        {mementos}
\bibfield{author}{\bibinfo{person}{Alice Thudt}, \bibinfo{person}{Dominikus Baur}, \bibinfo{person}{Samuel Huron}, {and} \bibinfo{person}{Sheelagh Carpendale}.} \bibinfo{year}{2016}\natexlab{}.
\newblock \showarticletitle{Visual Mementos: Reflecting Memories with Personal Data}.
\newblock \bibinfo{journal}{\emph{IEEE Transactions on Visualization and Computer Graphics}} \bibinfo{volume}{22}, \bibinfo{number}{1} (\bibinfo{year}{2016}), \bibinfo{pages}{369--378}.
\newblock
\urldef\tempurl%
\url{https://doi.org/10.1109/TVCG.2015.2467831}
\showDOI{\tempurl}


\bibitem[Thudt et~al\mbox{.}(2018)]%
        {self-reflection}
\bibfield{author}{\bibinfo{person}{Alice Thudt}, \bibinfo{person}{Uta Hinrichs}, \bibinfo{person}{Samuel Huron}, {and} \bibinfo{person}{Sheelagh Carpendale}.} \bibinfo{year}{2018}\natexlab{}.
\newblock \showarticletitle{Self-Reflection and Personal Physicalization Construction}. In \bibinfo{booktitle}{\emph{Proceedings of the 2018 CHI Conference on Human Factors in Computing Systems}} (<conf-loc>, <city>Montreal QC</city>, <country>Canada</country>, </conf-loc>) \emph{(\bibinfo{series}{CHI '18})}. \bibinfo{publisher}{Association for Computing Machinery}, \bibinfo{address}{New York, NY, USA}, \bibinfo{pages}{1–13}.
\newblock
\showISBNx{9781450356206}
\urldef\tempurl%
\url{https://doi.org/10.1145/3173574.3173728}
\showDOI{\tempurl}


\bibitem[Van~Garderen(2006)]%
        {math}
\bibfield{author}{\bibinfo{person}{Delinda Van~Garderen}.} \bibinfo{year}{2006}\natexlab{}.
\newblock \showarticletitle{Spatial Visualization, Visual Imagery, and Mathematical Problem Solving of Students With Varying Abilities}.
\newblock \bibinfo{journal}{\emph{Journal of Learning Disabilities}}  \bibinfo{volume}{39} (\bibinfo{date}{12} \bibinfo{year}{2006}), \bibinfo{pages}{496--506}.
\newblock
\urldef\tempurl%
\url{https://doi.org/10.1177/00222194060390060201}
\showDOI{\tempurl}


\bibitem[Watson et~al\mbox{.}(2019)]%
        {decision}
\bibfield{author}{\bibinfo{person}{Joanne Watson}, \bibinfo{person}{Hille Voss}, {and} \bibinfo{person}{Melissa~J. Bloomer}.} \bibinfo{year}{2019}\natexlab{}.
\newblock \showarticletitle{Placing the Preferences of People with Profound Intellectual and Multiple Disabilities At the Center of End-of-Life Decision Making Through Storytelling}.
\newblock \bibinfo{journal}{\emph{Research and Practice for Persons with Severe Disabilities}} \bibinfo{volume}{44}, \bibinfo{number}{4} (\bibinfo{year}{2019}), \bibinfo{pages}{267--279}.
\newblock
\urldef\tempurl%
\url{https://doi.org/10.1177/1540796919879701}
\showDOI{\tempurl}
\showeprint{https://doi.org/10.1177/1540796919879701}


\bibitem[While et~al\mbox{.}(2024)]%
        {While2024}
\bibfield{author}{\bibinfo{person}{Zack While}, \bibinfo{person}{Tanja Blascheck}, \bibinfo{person}{Yujie Gong}, \bibinfo{person}{Petra Isenberg}, {and} \bibinfo{person}{Ali Sarvghad}.} \bibinfo{year}{2024}\natexlab{}.
\newblock \showarticletitle{Glanceable Data Visualizations for Older Adults: Establishing Thresholds and Examining Disparities Between Age Groups}. In \bibinfo{booktitle}{\emph{Proceedings of the CHI Conference on Human Factors in Computing Systems}} \emph{(\bibinfo{series}{CHI ’24})}. \bibinfo{publisher}{ACM}.
\newblock
\urldef\tempurl%
\url{https://doi.org/10.1145/3613904.3642776}
\showDOI{\tempurl}


\bibitem[Wimer et~al\mbox{.}(2024)]%
        {wimer2024beyond}
\bibfield{author}{\bibinfo{person}{Brianna~Lynn Wimer}, \bibinfo{person}{Laura South}, \bibinfo{person}{Keke Wu}, \bibinfo{person}{Danielle~Albers Szafir}, \bibinfo{person}{Michelle~A Borkin}, {and} \bibinfo{person}{Ronald Metoyer}.} \bibinfo{year}{2024}\natexlab{}.
\newblock \showarticletitle{Beyond Vision Impairments: Redefining the Scope of Accessible Data Representations}.
\newblock \bibinfo{journal}{\emph{IEEE Transactions on Visualization and Computer Graphics}} (\bibinfo{year}{2024}).
\newblock


\bibitem[Wood et~al\mbox{.}(2024)]%
        {hdv}
\bibfield{author}{\bibinfo{person}{Rachel Wood}, \bibinfo{person}{Jinjuan~Heidi Feng}, {and} \bibinfo{person}{Jonathan Lazar}.} \bibinfo{year}{2024}\natexlab{}.
\newblock \showarticletitle{Health Data Visualization Literacy Skills of Young Adults with Down Syndrome and the Barriers to Inference-making}.
\newblock  \bibinfo{volume}{17}, \bibinfo{number}{1}, Article \bibinfo{articleno}{4} (\bibinfo{date}{mar} \bibinfo{year}{2024}), \bibinfo{numpages}{1}~pages.
\newblock
\showISSN{1936-7228}
\urldef\tempurl%
\url{https://doi.org/10.1145/3648621}
\showDOI{\tempurl}


\bibitem[Wu et~al\mbox{.}(2021)]%
        {chi21}
\bibfield{author}{\bibinfo{person}{Keke Wu}, \bibinfo{person}{Emma Petersen}, \bibinfo{person}{Tahmina Ahmad}, \bibinfo{person}{David Burlinson}, \bibinfo{person}{Shea Tanis}, {and} \bibinfo{person}{Danielle~Albers Szafir}.} \bibinfo{year}{2021}\natexlab{}.
\newblock \showarticletitle{Understanding Data Accessibility for People with Intellectual and Developmental Disabilities}. In \bibinfo{booktitle}{\emph{Proceedings of the 2021 CHI Conference on Human Factors in Computing Systems}} (Yokohama, Japan) \emph{(\bibinfo{series}{CHI '21})}. \bibinfo{publisher}{Association for Computing Machinery}, \bibinfo{address}{New York, NY, USA}.
\newblock
\showISBNx{9781450380966}
\urldef\tempurl%
\url{https://doi.org/10.1145/3411764.3445743}
\showDOI{\tempurl}


\bibitem[Wu and Szafir(2023)]%
        {vis23}
\bibfield{author}{\bibinfo{person}{Keke Wu} {and} \bibinfo{person}{Danielle~Albers Szafir}.} \bibinfo{year}{2023}\natexlab{}.
\newblock \showarticletitle{Empowering People with Intellectual and Developmental Disabilities through Cognitively Accessible Visualizations}. In \bibinfo{booktitle}{\emph{2023 IEEE Workshop on Visualization for Social Good (VIS4Good)}}. \bibinfo{publisher}{IEEE Computer Society}, \bibinfo{address}{Los Alamitos, CA, USA}.
\newblock
\urldef\tempurl%
\url{https://doi.org/10.1109/VIS4Good60218.2023.00007}
\showDOI{\tempurl}


\bibitem[Wu et~al\mbox{.}(2019)]%
        {viscomm}
\bibfield{author}{\bibinfo{person}{Keke Wu}, \bibinfo{person}{Shea Tanis}, {and} \bibinfo{person}{Danielle Albers~Szafir}.} \bibinfo{year}{2019}\natexlab{}.
\newblock \showarticletitle{Designing Communicative Visualization for People with Intellectual Developmental Disabilities}.
\newblock  (\bibinfo{date}{08} \bibinfo{year}{2019}).
\newblock
\urldef\tempurl%
\url{https://doi.org/10.31219/osf.io/zbjhr}
\showDOI{\tempurl}


\bibitem[Wu et~al\mbox{.}(2023)]%
        {chi23}
\bibfield{author}{\bibinfo{person}{Keke Wu}, \bibinfo{person}{Michelle~Ho Tran}, \bibinfo{person}{Emma Petersen}, \bibinfo{person}{Varsha Koushik}, {and} \bibinfo{person}{Danielle~Albers Szafir}.} \bibinfo{year}{2023}\natexlab{}.
\newblock \showarticletitle{Data, Data, Everywhere: Uncovering Everyday Data Experiences for People with Intellectual and Developmental Disabilities}. In \bibinfo{booktitle}{\emph{Proceedings of the 2023 CHI Conference on Human Factors in Computing Systems}} (Hamburg, Germany) \emph{(\bibinfo{series}{CHI '23})}. \bibinfo{publisher}{Association for Computing Machinery}, \bibinfo{address}{New York, NY, USA}.
\newblock
\showISBNx{9781450394215}
\urldef\tempurl%
\url{https://doi.org/10.1145/3544548.3581204}
\showDOI{\tempurl}


\bibitem[Young et~al\mbox{.}(2011)]%
        {sensorystory}
\bibfield{author}{\bibinfo{person}{Hannah Young}, \bibinfo{person}{Maggi Fenwick}, \bibinfo{person}{Loretto Lambe}, {and} \bibinfo{person}{James Hogg}.} \bibinfo{year}{2011}\natexlab{}.
\newblock \showarticletitle{Multi-sensory storytelling as an aid to assisting people with profound intellectual disabilities to cope with sensitive issues: A multiple research methods analysis of engagement and outcomes}.
\newblock \bibinfo{journal}{\emph{European Journal of Special Needs Education}}  \bibinfo{volume}{26} (\bibinfo{date}{05} \bibinfo{year}{2011}), \bibinfo{pages}{127--142}.
\newblock
\urldef\tempurl%
\url{https://doi.org/10.1080/08856257.2011.563603}
\showDOI{\tempurl}


\bibitem[Zdanovic et~al\mbox{.}(2022)]%
        {recall}
\bibfield{author}{\bibinfo{person}{Dominyk Zdanovic}, \bibinfo{person}{Tanja~Julie Lembcke}, {and} \bibinfo{person}{Toine Bogers}.} \bibinfo{year}{2022}\natexlab{}.
\newblock \showarticletitle{The Influence of Data Storytelling on the Ability to Recall Information}. In \bibinfo{booktitle}{\emph{Proceedings of the 2022 Conference on Human Information Interaction and Retrieval}} (Regensburg, Germany) \emph{(\bibinfo{series}{CHIIR '22})}. \bibinfo{publisher}{Association for Computing Machinery}, \bibinfo{address}{New York, NY, USA}, \bibinfo{pages}{67–77}.
\newblock
\showISBNx{9781450391863}
\urldef\tempurl%
\url{https://doi.org/10.1145/3498366.3505755}
\showDOI{\tempurl}


\end{thebibliography}

\end{document}